\documentclass[journal]{IEEEtran}
\usepackage{cite}
\usepackage{amsmath,amssymb,amsfonts}
\usepackage{algorithmic}
\usepackage{hyperref}
\usepackage{booktabs}
\usepackage{threeparttable}
\usepackage{subcaption}
\usepackage{makecell}
\usepackage{graphicx,color}
\usepackage{multirow}
\usepackage{textcomp}
\def\BibTeX{{\rm B\kern-.05em{\sc i\kern-.025em b}\kern-.08em
    T\kern-.1667em\lower.7ex\hbox{E}\kern-.125emX}}

\begin{document}

\title{Transmitter Noise Propagation in Millimeter-Wave and Sub-Terahertz: From Limits to Design Guidelines}

\author{Mahir~Burak~Usta,
Didem~Aydogan,
Evgenii~Vinogradov,
Mohammad~Shahmoradi,
Eduard~Alarc\'on,
Sergi~Abadal,
and~Korkut~Kaan~Tokgoz
\thanks{M. B. Usta, D. Aydogan, and K. K. Tokgoz are with the Faculty of Engineering and Natural Sciences, Sabanc\i\ University, Istanbul, Turkey (e-mail: mahirusta@sabanciuniv.edu).}
\thanks{E. Vinogradov, M. Shahmoradi, E. Alarc\'on, and S. Abadal are with the NaNoNetworking Center in Catalunya (N3Cat), Universitat Polit\`{e}cnica de Catalunya, Barcelona, Spain.}
\thanks{This work was supported by TUBITAK 121C073, Horizon Europe MSCA Post-Doctoral Fellowship (101062272-ENSPEC6G), ERC Starting Grant (101042080-WINC), and the I+D+i project titled BLOSSOMS (grant PID2024-158530OB-I00) funded by MICIU/AEI/10.13039/501100011033/ and by ERDF/EU.}}

\maketitle

\begin{abstract}
This paper presents a comprehensive link budget analysis for millimeter wave (mm-Wave) and sub-Terahertz (sub-THz) communication systems with primary focus on transmitter (TX) noise propagation, an often overlooked impairment that can dominate in scenarios where path loss is insufficient to suppress TX noise below receiver thermal and atmospheric molecular noise levels. Unlike traditional thermal noise limited analyses, this work demonstrates that TX noise is amplified by component noise figures that degrade significantly with frequency, rising from single digits to more than $15\,\mathrm{dB}$ in the sub-THz range. In the scenarios analyzed, this propagated TX noise reduces the achievable Signal-to-Noise Ratio (SNR) by approximately $15$ to $25\,\mathrm{dB}$ at short distances, creating fundamental SNR ceilings at ranges below about $10\,\mathrm{cm}$. We develop a systematic framework quantifying TX noise dominance conditions as functions of distance, frequency, and component parameters, revealing fundamental performance constraints for short-range next generation wireless systems. Our findings indicate that the TX noise figure should be as low as possible for short-range communication, and both TX noise and atmospheric molecular noise should be considered for medium- and long-range links.
\end{abstract}


%%%%%%%%%%%%%%%%%%%%%%%%%%%%%%%%%%%%%%%%%%%%%%%%%%%%%%%%%%%%%%
%%%%%%%%%%%%%%%%%%%%%%%%%%%%%%%%%%%%%%%%%%%%%%%%%%%%%%%%%%%%%%
%%%%%%%%%%%%%%%%%%%%section-start%%%%%%%%%%%%%%%%%%%%%%%%%%%
%%%%%%%%%%%%%%%%%%%%%%%%%%%%%%%%%%%%%%%%%%%%%%%%%%%%%%%%%%%%%%
%%%%%%%%%%%%%%%%%%%%%%%%%%%%%%%%%%%%%%%%%%%%%%%%%%%%%%%%%%%%%%
\section{INTRODUCTION}
The relentless demand for higher data rates has propelled wireless communication into the millimeter-wave (mm-Wave) and sub-terahertz (sub-THz) spectrum. The forthcoming sixth generation (6G) wireless communication technology will rely heavily on mm-Wave and sub-THz frequencies \cite{tripathi2021millimeter}. However, operating at these higher frequencies introduces multiple challenges beyond traditional free space path loss (FSPL) considerations \cite{Rikkinen,kurner2016thz}.

Existing research has extensively studied link budget models for mm-Wave and sub-THz communication systems, with analyses typically centering on path loss, antenna hardware, and a receiver (RX) noise floor based on thermal limits and the RX's noise figure \cite{speed_bump,Rikkinen}. These conventional approaches have enabled impressive demonstrations, including a 21 km link at 140 GHz \cite{wu201721}, 64 Gbps over 850 m at 240 GHz \cite{kallfass201564}, and recent achievements such as 84 Gbps over 1.26 km at 220 GHz \cite{liu2024high} and 200 Gbps over 4.6 km at 125 GHz \cite{wei2024demonstration}. At shorter distances, sub-THz implementations have achieved 80 Gb/s over 3 cm at 300 GHz \cite{lee2019} and 100 Gb/s over 1 m at 230 GHz \cite{rodriguez2019terahertz}. Similarly, comprehensive technology assessments for D-band (110–170 GHz) systems \cite{dore2020technology} focus on hardware limitations and architectural considerations but typically do not explicitly model transmitter (TX) noise propagation effects. 

Table~\ref{tab:literature_comparison} summarizes representative link budget analyses in mm-Wave/sub-THz literature. A limited number of analytical and simulation-based studies have incorporated atmospheric molecular absorption and emission noise in the THz band, notably in the context of nanoscale communications~\cite{jornet2011channel}; however, these works do not explicitly model the propagation of TX generated noise. Hardware impairment studies, such as \cite{Suzuki}, have shown that practical transmitters introduce nonidealities such as TX-induced noise, IQ imbalance, and nonlinear distortion that simplified link budget models typically do not account for. While prior work appropriately focuses on thermal noise for their analyzed distance ranges, the physics at shorter distances fundamentally differs. At the millimeters to meters distances investigated in this work, the small path loss means that TX noise arrives at the RX with minimal attenuation, potentially reaching levels comparable to the RX thermal noise floor. As 6G systems push towards higher frequencies and emerging short-range operational scenarios, it becomes critical to investigate whether these established models capture all relevant noise contributions or if overlooked sources like TX noise significantly impact system performance.

%%%%%table-start%%%%%%%
\begin{table*}[t]
\centering
\caption{Comparison of Link Budget Analyses in the Literature}
\label{tab:literature_comparison}
\begin{tabular}{p{2.0cm}p{1.1cm}p{1.3cm}p{1.0cm}p{1.0cm}p{1.0cm}p{1.0cm}p{1.0cm}p{1.0cm}p{1.0cm}}
\hline
Parameter & This Work & \cite{jornet2011channel} & \cite{lee2019} & \cite{rodriguez2019terahertz} & \cite{dore2020technology} & \cite{kallfass201564} & \cite{liu2024high} & \cite{wei2024demonstration} & \cite{wu201721} \\
\hline
Distance (m) & 0.001--1000 & $10^{-4}$--1 & 0.03 & 1 & 300 & 850 & 1260 & 4600 & 21000 \\
Freq. (GHz) & 30--450 & 100--10000 & 265.7 & 230 & 150 & 240 & 220 & 125--145 & 140 \\
Methodology & Sim. & Sim. & Exp. & Exp. & Sim. & Exp. & Exp. & Exp. & Exp. \\
TX Noise$^*$ & \checkmark & - & - & - & - & - & - & - & - \\
Atm. Noise$^*$ & \checkmark & \checkmark & - & - & - & - & - & - & - \\
\hline
\end{tabular}
\vspace{0.1cm}
\\
\footnotesize
$^*$These noise sources were not explicitly modeled in prior works, as they are typically negligible for the link distances and system assumptions considered. \\
\textit{Sim.: Simulation-based study; Exp.: Experimental study.}
\end{table*}
%%%%%table-start%%%%%%%

Despite the success of these demonstrations, most link budgets still do not explicitly model TX noise for short-range mm-Wave links, instead folding unmodeled impairments into a generic link margin. This work closes that gap by presenting a comprehensive link budget framework that treats propagated TX noise as a primary noise source rather than a margin term. We trace TX noise to its origins in component limitations and analyze its impact across mm-Wave and sub-THz frequencies for distances ranging from millimeters to kilometers. By turning an implicit margin into explicit noise modeling, our framework provides actionable guidance for next-generation system design.

The key contributions of this paper are as follows:
\begin{itemize}
    \item \textbf{TX Noise Propagation Framework:} We develop an explicit model for TX noise propagation in mm-Wave and sub-THz link budgets, tracing its impact from component level noise figures through wireless channel propagation to RX input, and derive a closed form analytical threshold that
    determines when propagated TX noise dominates performance.
    
    \item \textbf{Unified High Frequency Noise Framework:} We present a comprehensive link budget model that integrates thermal noise, atmospheric molecular noise, and propagated TX noise. In the scenarios analyzed propagated TX noise reduces the achievable Signal-to-Noise Ratio (SNR) by approximately $15$ to $25\,\mathrm{dB}$ at millimeter scale distances,
    demonstrating the critical impact of this often overlooked impairment.
\end{itemize}

The rest of the paper is organized as follows. Section II establishes the complete noise landscape for high-frequency wireless systems, providing physical origins and mathematical models for TX noise, atmospheric molecular noise, and propagation loss effects. Section III develops the TX noise dominance analytical framework, deriving the dominance criterion, and conducting parametric analysis across the design space. Section IV applies this framework through realistic case study scenarios. Section V presents the conclusions.

%%%%%figure-start%%%%%%%
\begin{figure*}[!t]
    \centering
    \includegraphics[width=1\textwidth]{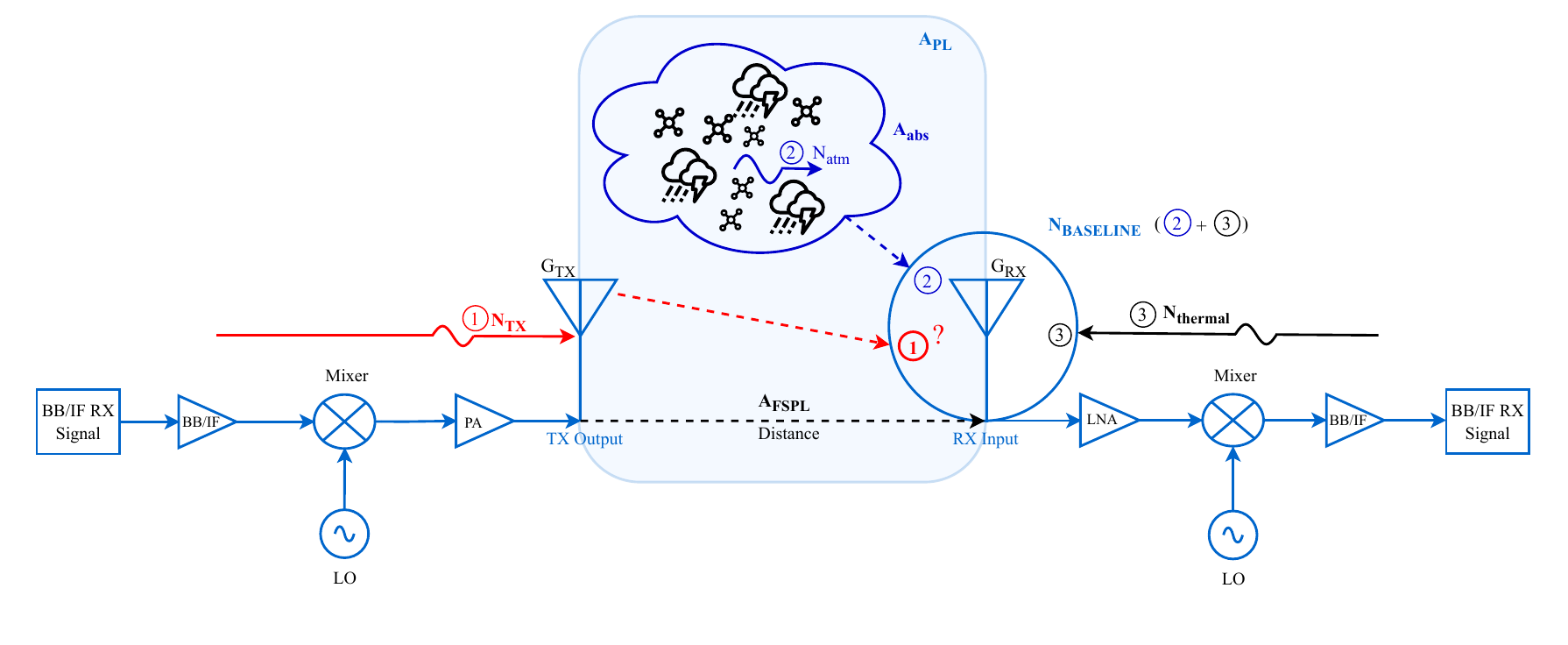}
    \caption{Conceptual diagram of the complete noise landscape. The total noise floor is shown to be a composite of three primary sources: (1) propagated TX noise ($N_{\text{TX}}$), (2) atmospheric molecular noise ($N_{\text{atm}}$), and (3) thermal noise ($N_{\text{thermal}}$). Here, $N_{\text{thermal}}$ represents the available thermal noise referenced to the RX input.
}
    \label{fig:system_figure}
\end{figure*}
%%%%%figure-end%%%%%%%

%%%%%%%%%%%%%%%%%%%%%%%%%%%%%%%%%%%%%%%%%%%%%%%%%%%%%%%%%%%%%%
%%%%%%%%%%%%%%%%%%%%%%%%%%%%%%%%%%%%%%%%%%%%%%%%%%%%%%%%%%%%%%
%%%%%%%%%%%%%%%%%%%%section-end%%%%%%%%%%%%%%%%%%%%%%%%%%%
%%%%%%%%%%%%%%%%%%%%%%%%%%%%%%%%%%%%%%%%%%%%%%%%%%%%%%%%%%%%%%
%%%%%%%%%%%%%%%%%%%%%%%%%%%%%%%%%%%%%%%%%%%%%%%%%%%%%%%%%%%%%%

%%%%%%%%%%%%%%%%%%%%%%%%%%%%%%%%%%%%%%%%%%%%%%%%%%%%%%%%%%%%%%
%%%%%%%%%%%%%%%%%%%%%%%%%%%%%%%%%%%%%%%%%%%%%%%%%%%%%%%%%%%%%%
%%%%%%%%%%%%%%%%%%%%section-start%%%%%%%%%%%%%%%%%%%%%%%%%%%
%%%%%%%%%%%%%%%%%%%%%%%%%%%%%%%%%%%%%%%%%%%%%%%%%%%%%%%%%%%%%%
%%%%%%%%%%%%%%%%%%%%%%%%%%%%%%%%%%%%%%%%%%%%%%%%%%%%%%%%%%%%%%
\section{COMPLETE NOISE LANDSCAPE}
\label{sec:section2}
In mm-Wave and sub-THz systems, the RX input noise is conventionally assumed to be dominated by the RX’s own thermal noise and noise figure. However, two additional noise sources can become significant: propagated TX noise and atmospheric molecular noise. Propagated TX noise can become the limiting factor in scenarios where path loss is insufficient to attenuate the TX noise floor below the RX thermal noise level. This occurs primarily at short ranges ($<$1~m), where the propagation loss is small. Atmospheric effects further degrade performance through frequency-selective attenuation and molecular emission noise, with the most severe degradation occurring in molecular absorption bands around 60~GHz and 180~GHz. This complete noise landscape is illustrated conceptually in Fig.~\ref{fig:system_figure}, which shows how TX noise propagates through the channel and combines with atmospheric and thermal contributions at the RX input. This view helps clarify how each noise source contributes to the total noise at the RX.

\subsection{TX Noise}
\label{sec:subsection2a}
Conventional SNR analysis usually assumes that the noise generated at the TX becomes negligible after propagation losses. At mm-Wave and sub-THz frequencies, however, the noise performance of active RF components degrades, and the TX chain contains only a limited number of gain stages. As a result, the additive noise generated at the TX can remain significant at short distances and therefore must be modeled explicitly. In this work, the propagated TX noise is determined by two dominant RF blocks in the TX chain (Figure~\ref{fig:system_figure}): the upconversion mixer with its immediate intermediate frequency (IF) or load stage, and the power amplifier (PA). Modern Complementary Metal-Oxide-Semiconductor (CMOS) and Silicon-Germanium (SiGe) RF implementations rarely provide standalone baseband (BB) or IF amplifier noise figure and gain at these frequencies. For this reason, the ``Mixer + BB or IF Amplifier'' entry in Table~\ref{tab:component_nf} represents the combined active frequency conversion block extracted from recent CMOS and SiGe designs. Local oscillator (LO) phase noise is not included in this analysis, since it primarily degrades error vector magnitude (EVM) and does not directly affect the additive thermal noise floor studied here.

\subsubsection{Amplifier Noise}
Amplifier noise critically impacts the overall TX noise performance at mm-Wave and sub-THz frequencies, where device noise mechanisms worsen and high gain becomes increasingly difficult to achieve. In the TX chain modeled in this work, the immediate BB or IF load stage is embedded inside the upconversion mixer core, and the PA is the only standalone amplification stage. Since direct noise figure measurements for PAs are rarely reported in the literature, most PA designs prioritize output power, efficiency, and linearity. PA noise figures in Table~\ref{tab:component_nf} are approximated from state-of-the-art low noise amplifier (LNA) baselines in the same technologies and frequency ranges, with an added 3 dB penalty to account for large signal operation.

For CMOS technology, the estimated PA noise figure rises from 7.8 dB in the 30--100 GHz band to 12.7 dB in the 200--500 GHz band~\cite{Wang_cmosPA_NF_lowband,Lee_cmosPA_NF_midband,Wang_cmosPA_NF_highband}. As a representative upper band reference, Wang \textit{et al.}\ report a 220 GHz LNA in 40 nm CMOS with a measured noise figure of 9.7 dB, which provides a realistic baseline for PA noise estimation in this range. CMOS PAs employing multiway power combining have demonstrated small signal gains exceeding 20 dB across these bands~\cite{Larie_cmosPA_gain_lowband,Tang_cmosPA_gain_midband,Yun_cmosPA_gain_highband}, but their intrinsic noise performance remains weaker than that of bipolar technologies. SiGe technology exhibits superior noise and gain characteristics at mm-Wave and sub-THz frequencies due to its higher transconductance and breakdown voltage, which support efficient large signal operation. The estimated SiGe PA noise figure ranges from 7.2 dB to 14.0 dB across the three bands considered~\cite{Ustundag_sigePA_NF_lowband,Maiwald_sigePA_NF_midband,Singh_sigePA_NF_highband}. High-gain SiGe designs further reduce the PA's effective noise contribution; for example, Bücher \textit{et al.}\ demonstrate a 300 GHz amplifier achieving 23 dB of small signal gain~\cite{Bucher_sigePA_gain_highband}, validating the ability of SiGe to sustain high gain and low noise at sub-THz frequencies.

\subsubsection{Mixer Noise}
The upconversion mixer stage introduces a clear divergence between CMOS and SiGe technologies at mm-Wave and sub-THz frequencies. In this work, the ``Mixer + BB or IF Amplifier'' block represents the combined active frequency conversion stage, including the embedded IF or BB load commonly co-designed with the mixer core. Because standalone TX mixer measurements above 100 GHz are scarce, the values in Table~\ref{tab:component_nf} are taken from closely related RX mixer first architectures~\cite{Memioglu_cmosMXR_BOTH_highband,Turkmen_sigeMXR_BOTH_highband} and from bidirectional D-band designs. This approach is justified because passive mixer topologies exhibit small signal reciprocity, and active Gilbert-cell mixers share nearly identical device level noise and gain behavior in upconversion and downconversion modes. As a result, well characterized RX mixers provide reliable proxies for estimating TX mixer noise performance at these frequencies.

SiGe implementations predominantly employ active Gilbert-cell topologies that deliver substantial conversion gain, ranging from 15 dB to 20.6 dB across the 30--500 GHz
range~\cite{Trotta_sigeMXR_BOTH_lowband,Testa_sigeMXR_BOTH_midband,Turkmen_sigeMXR_BOTH_highband}. 
This conversion gain is critical because it suppresses the noise contribution of the subsequent PA stage in the TX chain. 
Correspondingly, SiGe mixer noise figures remain low (11.2--13.2 dB) even as operating frequencies extend into the sub-THz region, reflecting the superior transconductance and high frequency capability of SiGe devices. CMOS mixers, by contrast, often rely on passive or resistive switching topologies in the upper mmWave and sub-THz bands where voltage headroom is limited. These designs exhibit conversion loss or low gain (from $-2$ dB to $+3$ dB)~\cite{Zhu_cmosMXR_BOTH_lowband,Lee_cmosMXR_BOTH_midband,Memioglu_cmosMXR_BOTH_highband}, which prevents significant suppression of downstream PA noise. Although CMOS mixer noise figures remain competitive in the lower and mid bands (approximately 12.4--16.0 dB), the combination of limited gain and increased passive loss at higher frequencies results in a higher effective transmitter noise floor, particularly in the 200--500 GHz band where device and switch resistance become dominant.

\subsubsection{Cascaded TX Noise Figure}
\begin{figure}[!t]
    \centerline{\includegraphics[width=1\columnwidth]{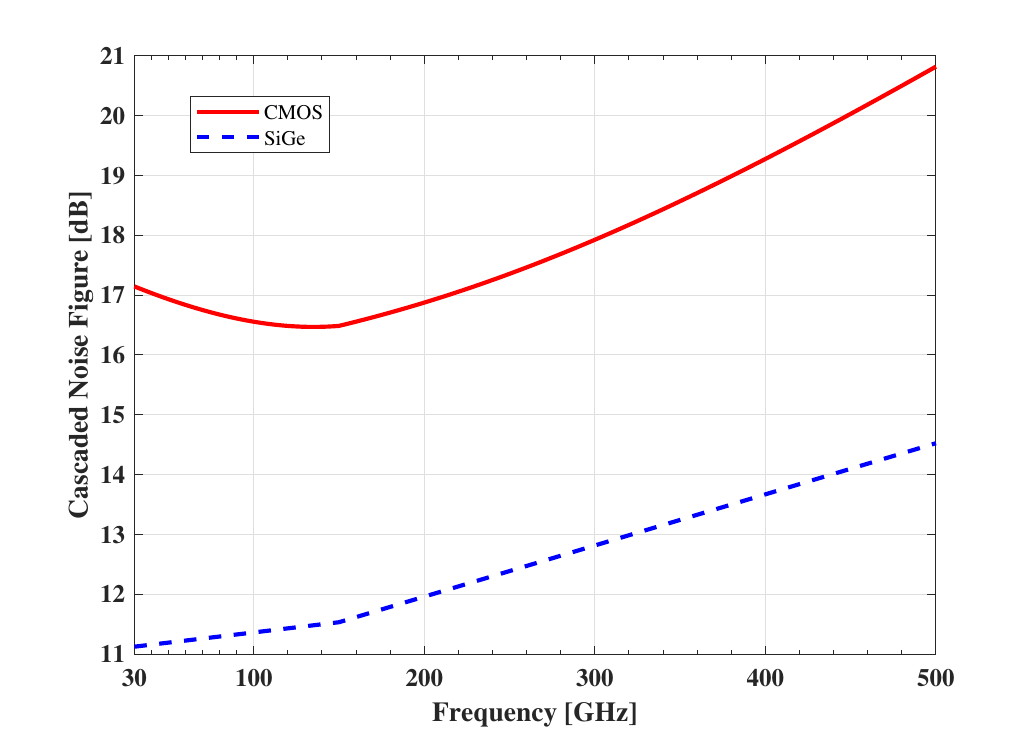}}
    \caption{Cascaded TX Chain Noise Figure ($F_{TX}$) vs. Frequency, derived from Table~\ref{tab:component_nf} using frequency interpolated component parameters and cascaded noise figure calculations.}
    \label{fig:cascaded_nf}
\end{figure}

The cumulative effect of these individual component limitations determines the total cascaded noise figure ($F_{TX}$) of the TX. It is calculated from the contributions of its components using the standard formula:
\begin{equation}
    F_{TX} = F_1 + \frac{F_2 - 1}{G_1} + \frac{F_3 - 1}{G_1 G_2} + \ldots,
    \label{equ:cascaded}
\end{equation}
where $F_i$ and $G_i$ are the noise figure and gain of the $i$-th component in the TX chain, respectively. As illustrated in Figure~\ref{fig:cascaded_nf}, this cascaded noise figure increases with frequency for both CMOS and SiGe implementations. Using frequency interpolated component values from Table~\ref{tab:component_nf}, CMOS chains exhibit cascaded noise figures ranging from approximately 17 dB at 30 GHz to 20.8 dB at 500 GHz, while SiGe implementations show improved performance from 11 dB to 14.5 dB across the same range.

The total TX output noise power spectral density ($N_{\text{TX}}$) is determined by the cascaded noise figure and the environmental temperature $T_{\text{env}}$:
\begin{equation}
N_{\text{TX}} = k \times T_{\text{env}} \times F_{\text{TX}},
\end{equation}
where $k$ is Boltzmann's constant and $T_{\text{env}}$ varies from 250 K to 310 K depending on altitude, humidity, and atmospheric conditions. This TX noise propagates through the wireless channel to the RX input; Section~\ref{sec:section3} develops the complete link budget framework including antenna gains and path loss effects.

%%%%%table-start%%%%%%%
\begin{table*}[!t]
\centering
\caption{Noise figure and gain characteristics of mm-Wave/sub-THz TX components used in the cascaded noise model}
\label{tab:component_nf}
\renewcommand{\arraystretch}{1.35}
\setlength{\tabcolsep}{4.2pt}
\scriptsize
\begin{threeparttable}
\begin{tabular}{@{}llcccccc@{}}
\toprule
\textbf{Component} & \textbf{Technology} &
\multicolumn{3}{c}{\textbf{Noise Figure [dB]}} &
\multicolumn{3}{c}{\textbf{Gain / Conv. Loss [dB]}} \\
\cmidrule(lr){3-5} \cmidrule(lr){6-8}
& & \textbf{30--100 GHz} & \textbf{100--200 GHz} & \textbf{200--500 GHz} 
& \textbf{30--100 GHz} & \textbf{100--200 GHz} & \textbf{200--500 GHz} \\ 
\midrule
%%%% PA %%%%
\multirow{2}{*}{\makecell[l]{Power\\Amplifier}} 
& CMOS 
& 7.8$^\dagger$~\cite{Wang_cmosPA_NF_lowband} 
& 7.9$^\dagger$~\cite{Lee_cmosPA_NF_midband} 
& 12.7$^\dagger$~\cite{Wang_cmosPA_NF_highband} 
& 15.4~\cite{Larie_cmosPA_gain_lowband}
& 22.5~\cite{Tang_cmosPA_gain_midband}
& 28.0~\cite{Yun_cmosPA_gain_highband} \\
& SiGe
& 7.2$^\dagger$~\cite{Ustundag_sigePA_NF_lowband}
& 9.0$^\dagger$~\cite{Maiwald_sigePA_NF_midband}
& 14.0$^\dagger$~\cite{Singh_sigePA_NF_highband}
& 17.0~\cite{Mortazavi_sigePA_gain_lowband}
& 20.0~\cite{Lin_sigePA_gain_midband}
& 23.0~\cite{Bucher_sigePA_gain_highband} \\
\midrule
%%%% MIXER + BB/IF %%%%
\multirow{2}{*}{\makecell[l]{Mixer +\\BB/IF Amplifier$^\ast$}} 
& CMOS 
& 16.0~\cite{Zhu_cmosMXR_BOTH_lowband}
& 14.5$^{\ddagger}$~\cite{Lee_cmosMXR_BOTH_midband}
& 18.0$^{\ast\ast}$~\cite{Memioglu_cmosMXR_BOTH_highband}
& $-$2.0~\cite{Zhu_cmosMXR_BOTH_lowband}
& $-$5.0$^{\ddagger}$~\cite{Lee_cmosMXR_BOTH_midband}
& 3.0$^{\ast\ast}$~\cite{Memioglu_cmosMXR_BOTH_highband} \\
& SiGe
& 11.2~\cite{Trotta_sigeMXR_BOTH_lowband}
& 11.5~\cite{Testa_sigeMXR_BOTH_midband}
& 13.2$^{\ast\ast}$~\cite{Turkmen_sigeMXR_BOTH_highband}
& 15.0~\cite{Trotta_sigeMXR_BOTH_lowband}
& 18.0~\cite{Testa_sigeMXR_BOTH_midband}
& 20.6$^{\ast\ast}$~\cite{Turkmen_sigeMXR_BOTH_highband} \\
\bottomrule
\end{tabular}
\vspace{0.3cm}
\begin{tablenotes}
\scriptsize
\item[$^\dagger$] PA noise figures are approximated from reported LNA values by adding a 3~dB penalty, a common rule-of-thumb when detailed PA NF data is not available.
\item[$^\ast$] "Gain / Conv. Loss" refers to mixer conversion gain, where negative values indicate conversion loss; depending on the reference, the reported value may include the immediate IF/load stage but not the full RX baseband chain.
\item[$^{\ast\ast}$] Due to the scarcity of standalone sub-THz TX mixer characterization, values are derived from the mixer first RX in \cite{Memioglu_cmosMXR_BOTH_highband} (CMOS) and the quadrature RX in \cite{Turkmen_sigeMXR_BOTH_highband} (SiGe), representing the achievable performance of the active core.
\item[$^{\ddagger}$] Noise figure is based on simulation results as measured data was not reported.
\end{tablenotes}
\end{threeparttable}
\end{table*}
%%%%%table-end%%%%%%%

%%%%%figure-start%%%%%%%
%%%%%figure-end%%%%%%%

\subsection{Atmospheric Molecular Noise}
\label{sec:subsection2b} 

At mm-Wave and sub-THz frequencies, molecular absorption and re-emission introduce additional noise contributions that critically impact signal propagation \cite{kokkoniemi2016discussion}. Unlike thermal noise, this molecular noise originates from electromagnetic interactions with atmospheric gases. The dominant absorber is water vapor (H$_2$O), which substantially raises the noise temperature near 183 GHz and 325 GHz \cite{slocum2013atmospheric}. Oxygen (O$_2$) is also a significant contributor around 60 GHz and in higher bands. Although molecular nitrogen and oxygen dominate the atmosphere's composition, it is primarily water vapor and, to a lesser extent, oxygen that govern electromagnetic propagation characteristics at these frequencies \cite{jornet2011channel}. Atmospheric molecular absorption simultaneously causes signal attenuation and noise generation \cite{box1986utilization}, with noise arising from molecular re-emission at the incident frequency, which increases as the distance increases.

Atmospheric absorption and transmittance can be modeled using standardized databases, such as the ITU-R P.676-10 model \cite{ITU} and the Atmospheric Model (AM) \cite{Paine}. In this work, the ITU model is employed, with simulations performed at sea level (1 m altitude), 35\textdegree C temperature, and 100\% relative humidity (corresponding to an absolute humidity of 39.6 g/m$^3$).

The atmospheric molecular noise is characterized by the channel's emissivity $\varepsilon$, which is derived from the transmissivity, where $d$ is the distance and $f$ is the frequency:
\begin{equation}
\varepsilon(d, f) = 1 - \tau(d, f)
\label{eq:emissivity}
\end{equation}
where $\tau(d, f)$ represents the atmospheric transmittance:
\begin{equation}
    \tau(d, f) = e^{-\kappa(f) \cdot d}
\end{equation}
where $\kappa(f)$ is the frequency-dependent absorption coefficient from \cite{ITU}.
The equivalent noise temperature due to atmospheric molecular absorption is given by:
\begin{equation}
T_{noise}(d, f) = T_0 \times \varepsilon(d, f) = T_0 \times (1 - \tau(d, f)).
\end{equation}
The corresponding atmospheric noise power spectral density is:
\begin{equation}
N_{atm} = k \times T_{noise}
\label{eq:atmospheric_npsd}
\end{equation}

\subsection{Propagation Loss and Baseline Noise}
\label{sec:subsection2c}
This subsection models the signal attenuation due to propagation and combines the non-TX noise contributions to establish a fundamental baseline noise floor. This baseline serves as the reference against which the impact of TX noise will be evaluated in our dominance analysis.

\subsubsection{Path Loss Components}
The propagation loss includes the free-space path loss (FSPL), calculated as:
\begin{equation}
A_{FSPL}(d, f) = 20 \times \log_{10}\left(\frac{4\pi df}{c}\right)
\end{equation}
where $c$ is the speed of light.

The atmospheric molecular absorption loss is calculated according to Beer-Lambert's Law as \cite{jornet2011channel}:
\begin{equation}
A_{abs}(d, f) = 10 \times \log_{10}\left(\frac{1}{\tau(d,f)}\right)
\end{equation}
where $\tau(d,f)$ is the atmospheric transmissivity defined in Eq.~\eqref{eq:emissivity}.

The total propagation loss combines both components:
\begin{equation}
A_{PL}(d, f) = A_{abs}(d, f) + A_{FSPL}(d, f) \quad [\text{dB}]
\end{equation}

\begin{figure}[!t]
   \centerline{\includegraphics[width=\columnwidth]{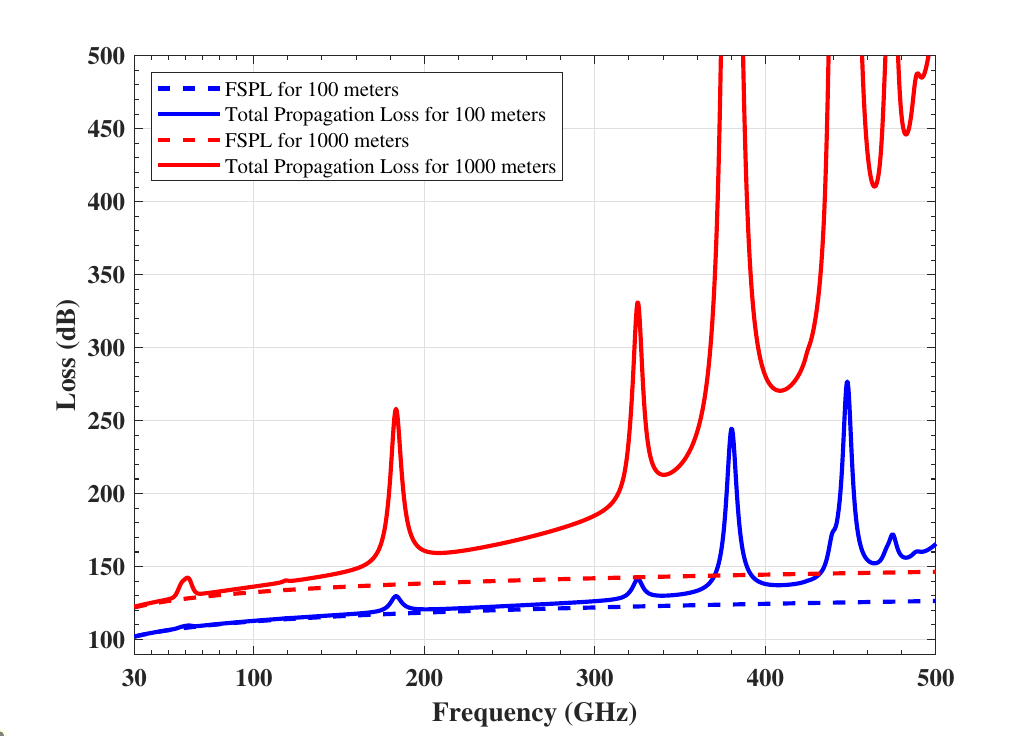}}
    \caption{FSPL and total propagation loss (FSPL + atmospheric molecular absorption) versus frequency for 100m and 1000m link distances. Atmospheric conditions: 35°C, 1011.9 mbar, 39.6 g/m³ water vapor density.}
    \label{fig:Propagation_Loss}
\end{figure}

Figure~\ref{fig:Propagation_Loss} illustrates the total propagation loss and FSPL for link distances of 100 meters and 1000 meters up to 500 GHz. Significant attenuation peaks are observed at approximately 60 GHz (oxygen absorption), 180 GHz, and 320 GHz (water vapor absorption). At a 1000-meter link distance, the 180 GHz band experiences more than 100 dB of atmospheric molecular absorption loss, compared to approximately 15 dB for a 100-meter link distance at the same frequency. Similar effects occur around 320 GHz, where a 1000-meter link experiences over 180 dB of atmospheric absorption loss, while a 100-meter link experiences around 20 dB.

%%%%%figure-start%%%%%%%
%%%%%figure-end%%%%%%%

\subsubsection{Baseline Noise Floor}
The fundamental thermal noise power spectral density is:
\begin{equation}
N_{\text{thermal}} = k \times T_{\text{env}}.
\end{equation}

The baseline noise floor at the RX input combines thermal and atmospheric molecular noise:
\begin{equation}
\label{eq:N_baseline}
N_{\text{Baseline}}(f) = N_{\text{thermal}} + N_{\text{atm}}(f).
\end{equation}

This baseline represents the noise floor of a system with a hypothetically noiseless TX and serves as the reference for evaluating TX noise dominance in Section~\ref{sec:section3}.

%%%%%%%%%%%%%%%%%%%%%%%%%%%%%%%%%%%%%%%%%%%%%%%%%%%%%%%%%%%%%%
%%%%%%%%%%%%%%%%%%%%%%%%%%%%%%%%%%%%%%%%%%%%%%%%%%%%%%%%%%%%%%
%%%%%%%%%%%%%%%%%%%%section-end%%%%%%%%%%%%%%%%%%%%%%%%%%%
%%%%%%%%%%%%%%%%%%%%%%%%%%%%%%%%%%%%%%%%%%%%%%%%%%%%%%%%%%%%%%
%%%%%%%%%%%%%%%%%%%%%%%%%%%%%%%%%%%%%%%%%%%%%%%%%%%%%%%%%%%%%%

%%%%%%%%%%%%%%%%%%%%%%%%%%%%%%%%%%%%%%%%%%%%%%%%%%%%%%%%%%%%%%
%%%%%%%%%%%%%%%%%%%%%%%%%%%%%%%%%%%%%%%%%%%%%%%%%%%%%%%%%%%%%%
%%%%%%%%%%%%%%%%%%%%section-start%%%%%%%%%%%%%%%%%%%%%%%%%%%
%%%%%%%%%%%%%%%%%%%%%%%%%%%%%%%%%%%%%%%%%%%%%%%%%%%%%%%%%%%%%%
%%%%%%%%%%%%%%%%%%%%%%%%%%%%%%%%%%%%%%%%%%%%%%%%%%%%%%%%%%%%%%
\section{TX NOISE DOMINANCE}
\label{sec:section3}
Having established the components of the complete noise landscape, this section derives the analytical conditions under which propagated TX noise becomes the dominant performance limiter.

\subsection{Derivation of the Dominance Criterion}
\label{sec:subsection3a}
We begin by expressing the TX noise power spectral density at the RX input, $N_{\text{TXtoRX}}$, where the TX noise propagates through the wireless channel and is affected by antenna gains and path loss:
\begin{equation}
    \label{eq:N_tx_at_rx_linear_new}
    N_{\text{TXtoRX}} = N_{\text{TX}} \times G_{\text{TX,lin}} \times G_{\text{RX,lin}} \times L_{\text{path,lin}}^{-1}
\end{equation}
where $G_{\text{TX,lin}}$ and $G_{\text{RX,lin}}$ are the TX and RX antenna gains in linear scale, and $L_{\text{path,lin}}$ denotes the linear-scale propagation loss corresponding to $A_{\text{PL}}(d,f)=A_{\text{FSPL}}(d,f)+A_{\text{abs}}(d,f)$ defined in Section~II-B.

\subsubsection{SNR Degradation-Based Threshold Derivation}
Rather than adopting an arbitrary engineering threshold, we derive the criterion based on acceptable SNR degradation. To do so, we first establish the conceptual SNR states. Let $P_{\text{signal}}$ represent the general received signal power. The baseline SNR, which we denote as $\text{SNR}_{\text{baseline}}$, represents the system's performance without considering TX noise:
\begin{equation}
\label{eq:snr_baseline}
\text{SNR}_{\text{baseline}} = \frac{P_{\text{signal}}}{N_{\text{baseline}}}
\end{equation}

When the propagated TX noise, $N_{\text{TXtoRX}}$, is included, the resulting SNR, which we denote as $\text{SNR}_{\text{TXtoRX}}$, becomes:
\begin{equation}
\label{eq:snr_with_tx}
\text{SNR}_{\text{TXtoRX}} = \frac{P_{\text{signal}}}{N_{\text{baseline}} + N_{\text{TXtoRX}}}
\end{equation}

From these definitions, SNR degradation in dB is:
\begin{equation}
\label{eq:snr_degradation_step1}
\begin{split}
\Delta\text{SNR}[\text{dB}] &= 10\log_{10}\left(\frac{\text{SNR}_{\text{baseline}}}{\text{SNR}_{\text{TXtoRX}}}\right) \\
&= 10\log_{10}\left(\frac{P_{\text{signal}}/N_{\text{baseline}}}{P_{\text{signal}}/(N_{\text{baseline}} + N_{\text{TXtoRX}})}\right)
\end{split}
\end{equation}

Simplifying by canceling $P_{\text{signal}}$:
\begin{equation}
\label{eq:snr_degradation}
\Delta\text{SNR}[\text{dB}] = 10\log_{10}\left(1 + \frac{N_{\text{TXtoRX}}}{N_{\text{baseline}}}\right)
\end{equation}

For a 1~dB maximum acceptable degradation, representing the point where TX noise begins to impact link budget margins, a critical loss in mm-Wave and sub-THz systems where margins are already severely constrained, we set:
\begin{equation}
\label{eq:1db_condition}
10\log_{10}\left(1 + \frac{N_{\text{TXtoRX}}}{N_{\text{baseline}}}\right) = 1
\end{equation}

Solving for the threshold ratio:
\begin{equation}
\label{eq:threshold_derivation}
1 + \frac{N_{\text{TXtoRX}}}{N_{\text{baseline}}} = 10^{1/10} = 1.259
\end{equation}

Therefore:
\begin{equation}
\label{eq:dominance_threshold}
\frac{N_{\text{TXtoRX}}}{N_{\text{baseline}}} = 0.259 \approx 0.26
\end{equation}

This establishes our dominance criterion:
\begin{equation}
\label{eq:dominance_inequality}
N_{\text{TXtoRX}}(f) \geq 0.26 \cdot N_{\text{baseline}}(f).
\end{equation}

\subsubsection{Path Loss Threshold Derivation}
To derive the path loss threshold, we substitute equation \eqref{eq:N_tx_at_rx_linear_new} into \eqref{eq:dominance_inequality}:
\begin{equation}
N_{\text{TX}} \times G_{\text{TX,lin}} \times G_{\text{RX,lin}} \times L_{\text{path,lin}}^{-1} \geq 0.26 \cdot N_{\text{baseline}}(f).
\end{equation}

Converting to the decibel domain using $X[\text{dB}] = 10\log_{10}(X)$ and $10\log_{10}(L_{\text{path,lin}}^{-1}) = -A_{\text{PL}}[\text{dB}]$:
\begin{equation}
\begin{split}
&N_{\text{TX}}[\text{dB}] + G_{\text{TX}}[\text{dBi}] + G_{\text{RX}}[\text{dBi}] - A_{\text{PL}}[\text{dB}] \\
&\qquad \geq N_{\text{baseline}}(f)[\text{dB}] + 10\log_{10}(0.26)
\end{split}
\end{equation}

Since $10\log_{10}(0.26) = -5.9$~dB, rearranging yields the dominance threshold:
\begin{equation}
\label{eq:PL_TH}
\begin{split}
A_{\text{PL,th}}(f) = &N_{\text{TX}}[\text{dB}] + G_{\text{TX}}[\text{dBi}] + G_{\text{RX}}[\text{dBi}] \\
&- N_{\text{baseline}}(f)[\text{dB}] + 5.9\text{ dB}.
\end{split}
\end{equation}

When the link's actual path loss is less than this threshold ($A_{\text{PL}} < A_{\text{PL,th}}$), TX noise causes more than 1 dB SNR degradation and therefore dominates; conversely, when $A_{\text{PL}} > A_{\text{PL,th}}$, TX noise contribution remains below 1 dB and the link is limited by thermal and atmospheric noise. For practical engineering checks, this can be expressed in link budget format:
\begin{equation}
\label{eq:link_budget_form}
N_{\text{TX}}[\text{dB}] + G_{\text{TX}}[\text{dBi}] + G_{\text{RX}}[\text{dBi}] - A_{\text{PL}}[\text{dB}] \geq -5.9 \text{ dB}.
\end{equation}

The derived framework establishes multiple severity levels beyond the 1~dB significance threshold. Notably, 3~dB degradation ($N_{\text{TXtoRX}} = N_{\text{baseline}}$) represents a critical threshold where the total noise power at the RX is doubled, often necessitating a step down to lower-order modulation schemes to maintain link integrity.

This analytical framework assumes far-field propagation conditions. For the antenna apertures used in this work, the Fraunhofer distance decreases rapidly with frequency and remains below the evaluated link distances across most of the 30--500~GHz band. Even at the lowest frequencies, where the shortest separations approach the radiating near-field boundary, the far-field model offers a conservative estimate of TX noise coupling. The elevated $F_{\text{TX}}$ values characteristic of mm-Wave and sub-THz components (Table~\ref{tab:component_nf}) make this approach essential for identifying when traditional thermal noise limited link budget assumptions break down. The complete significance level classification is summarized in Table~\ref{tab:threshold_tiers}.

%%%%%table-start%%%%%%%
\begin{table*}[t]
\centering
\caption{TX Noise Significance Thresholds and Corresponding Dominance Criteria}
\label{tab:threshold_tiers}
\begin{tabular}{c c c l}
\hline
\hline
\textbf{$\Delta$SNR} & 
\textbf{Ratio} $\left(\frac{N_{\text{TXtoRX}}}{N_{\text{baseline}}}\right)$ &
\textbf{Threshold} [dB] &
\textbf{Practical Impact} \\
\hline
$<1$ dB & $<0.26$ & $<-5.9$ & Negligible performance impact \\
$1$ dB & $0.26$ & $-5.9$ & Onset of TX noise dominance \\
$1$--$3$ dB & $0.26$--$1$ & $-5.9$ to $0$ & Link margin reduction, TX NF optimization needed \\
$3$ dB & $\approx 1$ & $0$ & Noise power doubled; modulation constraints \\
$3$--$5$ dB & $1$--$2.2$ & $0$ to $3.4$ & Severe performance degradation \\
$>5$ dB & $>2.2$ & $>3.4$ & System architecture changes required \\
\hline
\hline
\multicolumn{4}{l}{\scriptsize{Note: The Threshold column refers to the right-hand side of Eq. (\ref{eq:link_budget_form}), calculated as $10\log_{10}(10^{\Delta \text{SNR}/10}-1)$.}}
\end{tabular}
\end{table*}
%%%%%table-end%%%%%%%

\subsection{Parametric Analysis and Design Guidelines}
\label{sec:subsection3c}
To demonstrate the practical utility of the dominance criterion derived in Section III-A, we now explore the parameter space governing TX noise significance. This analysis reveals critical design insights for mm-Wave and sub-THz system development.

\subsubsection{Parametric Analysis}
\label{sec:parametric_sub}
To reveal the design space constraints imposed by TX noise, we systematically sweep key system parameters: TX noise figure, operating frequency, link distance, and total path loss. This parametric exploration identifies critical performance thresholds and quantifies trade offs between hardware quality and link geometry.

\begin{figure*}[!t]
    \centering
    \begin{subfigure}[t]{0.48\textwidth}
        \centering
        \includegraphics[width=\textwidth]{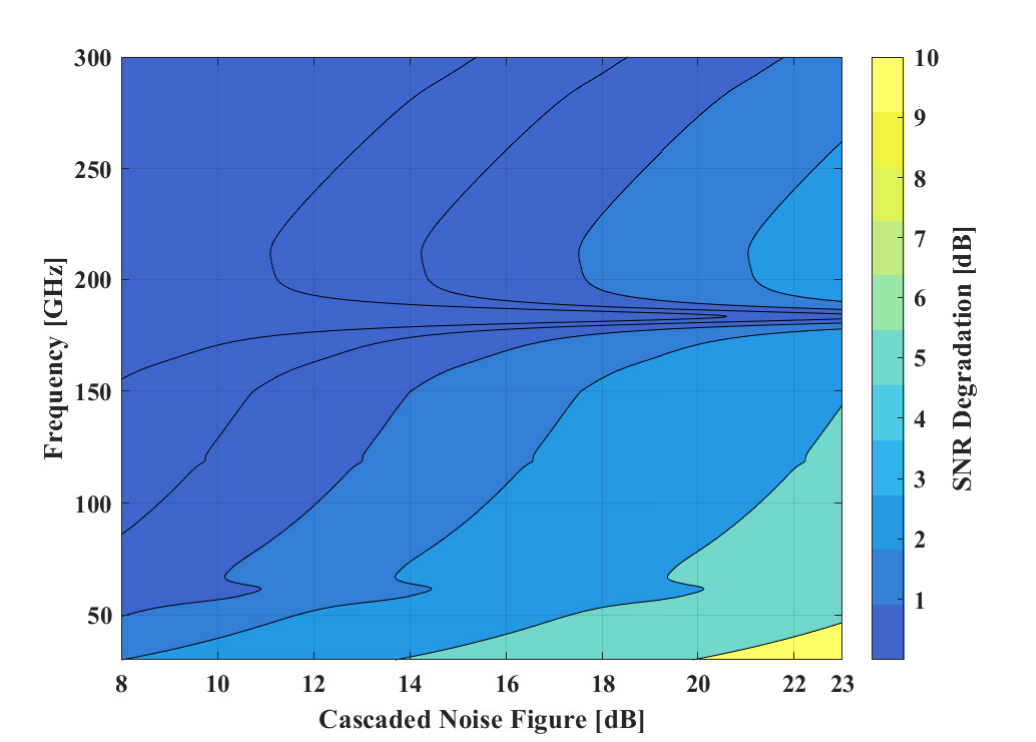} 
        \caption{Impact of TX hardware quality ($F_{TX}$) and operating frequency on SNR degradation for a fixed 100m link.}
        \label{fig:frequency_swept}
    \end{subfigure}
    \hfill
    \begin{subfigure}[t]{0.48\textwidth}
        \centering
        \includegraphics[width=\textwidth]{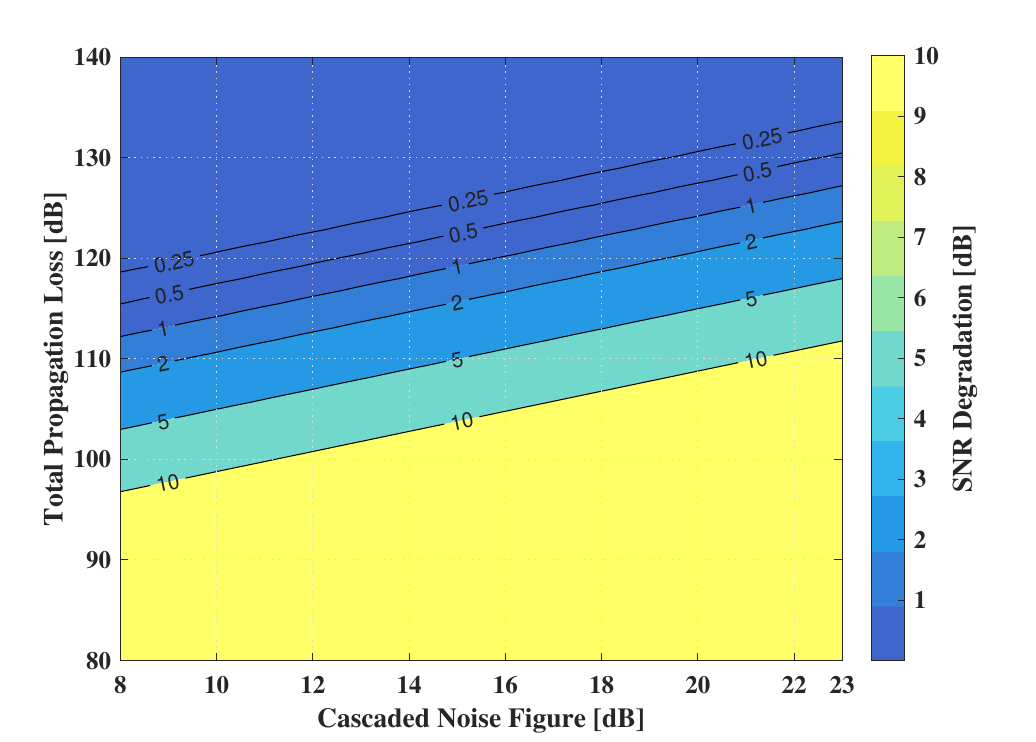}
        \caption{Design chart illustrating the trade-off between total path loss and TX noise figure ($F_{TX}$) for a link operating at 250~GHz.}
        \label{fig:pl_swept}
    \end{subfigure}
    \caption{Parametric analysis of TX noise impact. Plot (a) shows the frequency-dependent hardware requirements for a fixed link, while plot (b) provides a direct design tool for trading off link budget against hardware quality at a fixed frequency.}
    \label{fig:colormap}
\end{figure*}

Figure~\ref{fig:colormap} characterizes the transition regions where TX noise begins to influence system performance. Figure~\ref{fig:frequency_swept} shows the SNR degradation as a function of the cascaded TX noise figure ($8$--$23\,\mathrm{dB}$) and the operating frequency ($30$--$300\,\mathrm{GHz}$) for a fixed $100\,\mathrm{m}$ link. The colormap reveals that for $F_{\mathrm{TX}}$ below approximately $12\,\mathrm{dB}$, the degradation remains below about $2\,\mathrm{dB}$ for frequencies above $100\,\mathrm{GHz}$, whereas the lower frequency region ($30$--$70\,\mathrm{GHz}$) already exhibits $2$--$4\,\mathrm{dB}$ penalties. As $F_{\mathrm{TX}}$ increases beyond $15\,\mathrm{dB}$, the degradation becomes more pronounced, particularly below $80\,\mathrm{GHz}$ where losses reach $4$--$6\,\mathrm{dB}$. The most severe degradation occurs when the cascaded TX noise figure exceeds approximately $20\,\mathrm{dB}$ and the operating frequency falls below $50\,\mathrm{GHz}$, where the SNR penalty reaches the $10\,\mathrm{dB}$ level. The contour distortions near $60\,\mathrm{GHz}$ and above $180\,\mathrm{GHz}$ arise from atmospheric molecular absorption, which induces frequency selective changes in the baseline noise floor.

Figure~\ref{fig:pl_swept} illustrates the trade-off between total propagation loss and
the cascaded TX noise figure at $250\,\mathrm{GHz}$. The colormap quantifies the resulting
SNR degradation, with dark blue regions indicating negligible TX noise impact
($<\!1\,\mathrm{dB}$) and yellow regions representing severe penalties approaching
$10\,\mathrm{dB}$. The overlaid contour lines labeled $0.25$, $0.5$, $1$, $2$, $5$, and $10$
dB denote constant SNR degradation levels. For example, achieving less than $1\,\mathrm{dB}$ degradation at a total propagation loss of $115\,\mathrm{dB}$ requires a cascaded TX noise figure better than approximately $12\,\mathrm{dB}$. Conversely, if the TX exhibits a higher noise figure of $18\,\mathrm{dB}$, the system must operate with a total path loss above $130\,\mathrm{dB}$ to maintain the same $1\,\mathrm{dB}$ threshold.

\vspace{5pt}
\noindent \textbf{\textit{Design Insight:}} In the TX noise limited regime, high quality TX hardware (low $F_{\text{TX}}$) enables reliable operation at shorter distances, where the path loss is insufficient to attenuate propagated TX noise. In contrast, noisier TX chains require operation in higher loss regimes (i.e., longer distances or higher absorption) so that the channel sufficiently suppresses the TX noise before it reaches the RX.

%%%%%figure-start%%%%%%%
%%%%%figure-end%%%%%%%

%%%%%%%%%%%%%%%%%%%%%%%%%%%%%%%%%%%%%%%%%%%%%%%%%%%%%%%%%%%%%%
%%%%%%%%%%%%%%%%%%%%%%%%%%%%%%%%%%%%%%%%%%%%%%%%%%%%%%%%%%%%%%
%%%%%%%%%%%%%%%%%%%%section-end%%%%%%%%%%%%%%%%%%%%%%%%%%%
%%%%%%%%%%%%%%%%%%%%%%%%%%%%%%%%%%%%%%%%%%%%%%%%%%%%%%%%%%%%%%
%%%%%%%%%%%%%%%%%%%%%%%%%%%%%%%%%%%%%%%%%%%%%%%%%%%%%%%%%%%%%%

%%%%%%%%%%%%%%%%%%%%%%%%%%%%%%%%%%%%%%%%%%%%%%%%%%%%%%%%%%%%%%
%%%%%%%%%%%%%%%%%%%%%%%%%%%%%%%%%%%%%%%%%%%%%%%%%%%%%%%%%%%%%%
%%%%%%%%%%%%%%%%%%%%section-start%%%%%%%%%%%%%%%%%%%%%%%%%%%
%%%%%%%%%%%%%%%%%%%%%%%%%%%%%%%%%%%%%%%%%%%%%%%%%%%%%%%%%%%%%%
%%%%%%%%%%%%%%%%%%%%%%%%%%%%%%%%%%%%%%%%%%%%%%%%%%%%%%%%%%%%%%
\section{System Model and Implementation Assumptions}
\label{sec:system_model}

This section specifies the system implementation models, environmental parameters, and performance metrics used throughout the evaluation. These assumptions ground the theoretical analysis in practical, hardware-aware scenarios and define the baseline configurations for the subsequent case studies.

\subsection{Implementation Models and Parameters}
\label{sec:implementation_models}
To ground the preceding theoretical analysis in practical scenarios, this section details the specific implementation models, environmental parameters, and performance metrics used for the case studies that follow.

%%%%%figure-start%%%%%%%
\begin{figure}[!t]
    \centerline{\includegraphics[width=\columnwidth]{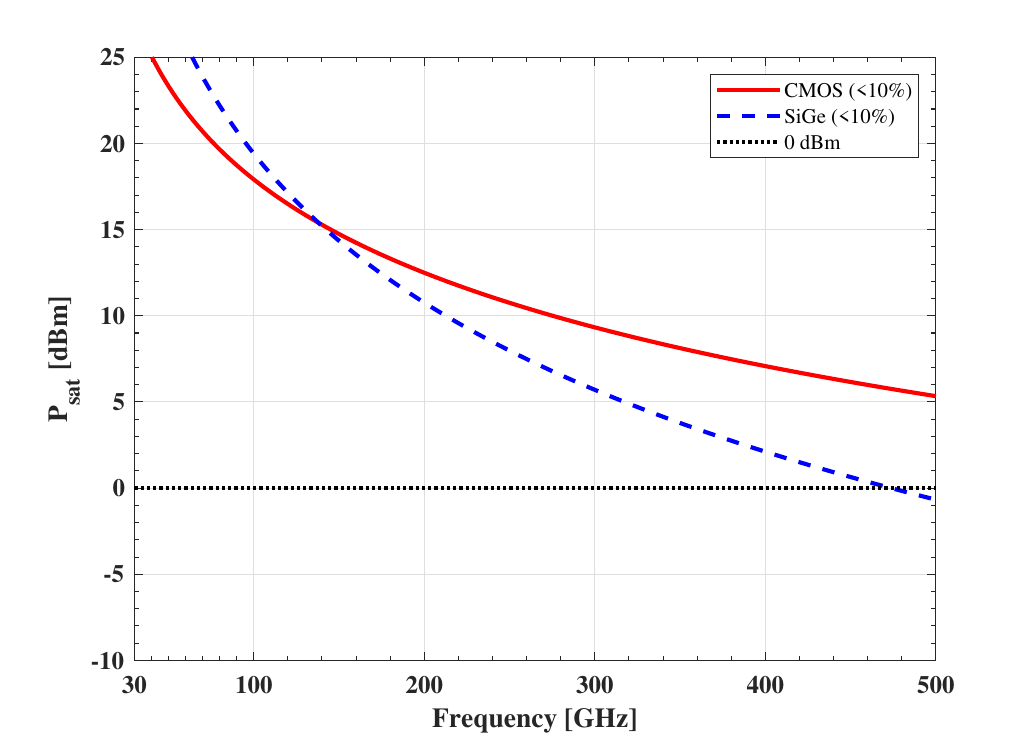}}
    \caption{Saturated output power ($P_{\text{sat}}$) models for CMOS and SiGe technologies versus frequency, derived from the ETH Zurich PA survey~\cite{eth_pa_survey}. }
\label{fig:eth_validation}
\end{figure}
%%%%%figure-end%%%%%%%

\subsubsection{Frequency-Dependent Transmit Power}
Real mm-Wave and sub-THz systems exhibit frequency-dependent power capabilities due to fundamental semiconductor limitations. We employ a CMOS-based power model derived from the ETH Zurich PA survey~\cite{eth_pa_survey}, which represent trendlines for amplifiers with a power added efficiency below 10\%. For CMOS technology, the model is:
\begin{equation}
P_{TX}(f) = 53.902 - 7.815 \log(f_{GHz}) \quad \text{[dBm]}
\label{equ:ptx}
\end{equation}

These models, illustrated in Figure~\ref{fig:eth_validation}, reveal the critical performance trade-off between these two key silicon based technologies (SiGe and CMOS). SiGe offers superior saturated output power ($P_{\text{sat}}$) at lower mm-Wave frequencies, starting at 25~dBm at 30~GHz. In contrast, the model suggests CMOS may have an advantage at higher frequencies, with a distinct crossover point occurring at approximately 140~GHz. For instance, at 300~GHz, the SiGe model delivers approximately 6~dBm, while the CMOS model provides 9~dBm. However, this high-frequency trend should be treated with caution, as it is an extrapolation based on fewer available data points for both technologies. This technology specific analysis is crucial, as the choice between CMOS and SiGe fundamentally alters the link budget at different operating frequencies.

%%%%%table-start%%%%%%%
\begin{table}[!t]
\centering
\caption{Environmental Conditions}
\label{tab:environmental_params}
\footnotesize
\begin{tabular}{@{}lccc@{}}
\toprule
\textbf{Condition} & \textbf{Temp (°C)} & \textbf{Pressure (Pa)} & \textbf{Humidity (g/m³)} \\
\midrule
Hot  & 35 & 101190 & 39.6 \\
Moderate & 15 & 101325 & 12.8 \\
Cold/Dry & -5 & 102100 & 3.4 \\
\bottomrule
\end{tabular}
\vspace{0.3cm}  % Add some vertical space
\end{table}
%%%%%table-end%%%%%%%

\subsubsection{Bandwidth Allocation}
The communication bandwidth ($\Delta f$) is modeled as a fraction of the carrier frequency ($f_c$), which is representative of ultra-wideband systems envisioned for 6G. For this work, a 25\% fractional bandwidth is used:
\begin{equation}
\Delta f = 0.25 \times f_c
\end{equation}
This assumption is aligned with the first standardization efforts for high-rate sub-THz communication. The IEEE 802.15.3d standard, for instance, defines a channel plan that includes a single channel with a massive 69.12 GHz bandwidth operating in the 252--322 GHz range\cite{petrov2020ieee}. A system utilizing this channel would have a fractional bandwidth of approximately 24\%, justifying the 25\% value used in our link budget model.

\subsubsection{Antenna Gain Models}
Antenna gains are assigned based on commercially achievable values for different link categories, reflecting practical aperture size constraints and deployment scenarios:

\begin{itemize}
\item \textbf{On-chip/Short Range ($<$1m):} 0 dBi - Isotropic or low-gain integrated antennas
\item \textbf{Medium Range (100m):} 40 dBi - Moderate aperture antennas for backhaul and infrastructure links  
\item \textbf{Long Range (1000m):} 56 dBi - High-gain parabolic reflectors or large aperture arrays for point-to-point communications
\end{itemize}

These gain values are representative of commercially available high-gain Cassegrain antennas for the mm-Wave and sub-THz spectrum, which offer gains in the 24-58 dBi range depending on operational frequency and reflector size~\cite{eravant_25dbi,eravant_40dbi,eravant_50dbi,eravant_53dbi,eravant_58dbi}. The symmetric antenna assumption ($G_{TX} = G_{RX}$) reflects typical point-to-point link deployments. Table~\ref{tab:system_params} summarizes the key system parameters used across the short-, medium-, and long-range case studies.

\subsection{Link Budget and Performance Metrics}
This subsection formalizes the link budget expressions and performance metrics used consistently across all case studies, based on the system models defined in the previous subsection. The received signal power, incorporating the frequency dependent models established above, is given by:
\begin{equation}
P_{RX}(f,d) = P_{TX}(f) - A_{PL}(d,f) + G_{TX}(d) + G_{RX}(d)
\label{eq:prx_definition}
\end{equation}

The total noise power at the RX input is calculated by integrating the baseline and propagated TX noise contributions over bandwidth $\Delta f$:
\begin{equation}
    \label{eq:total_noise_floor}
    P_{Noise, total} = \int_{\Delta f} \left( N_{Baseline}(f) + N_{TXtoRX}(f) \right) df
\end{equation}

The SNR at the RX follows from the ratio of signal to total noise power:
\begin{equation}
    \label{eq:snr_final}
    \text{SNR} = \frac{P_{RX}}{ P_{Noise, total} }
\end{equation}

Channel capacity is determined using the Shannon-Hartley theorem:
\begin{equation}
    \label{eq:capacity_final}
    C = \Delta f \times \log_2\left(1 + \text{SNR}\right)
\end{equation}

%%%%%table-start%%%%%%%
\begin{table}[!t]
\centering
\caption{System Implementation Parameters}
\label{tab:system_params}
\footnotesize
\setlength{\tabcolsep}{3pt}
\begin{tabular}{@{}lccc@{}}
\toprule
\textbf{Parameter} & \textbf{Short} & \textbf{Medium} & \textbf{Long} \\
\midrule
Distance Range & 1 mm -- 1 m & 100 m & 1 km \\
TX Power & 0 dBm & Eq.~\eqref{equ:ptx} & Eq.~\eqref{equ:ptx} \\
Antenna Gain (each) & 0 dBi & 40 dBi & 56 dBi \\
TX NF & \multicolumn{3}{c}{Fig.~\ref{fig:cascaded_nf} (freq.-dependent)} \\
Bandwidth & \multicolumn{3}{c}{$0.25 \times f_c$} \\
Frequency Range & \multicolumn{3}{c}{30 -- 500 GHz} \\
\bottomrule
\end{tabular}
\end{table}
%%%%%table-end%%%%%%%

%%%%%%%%%%%%%%%%%%%%%%%%%%%%%%%%%%%%%%%%%%%%%%%%%%%%%%%%%%%%%%
%%%%%%%%%%%%%%%%%%%%%%%%%%%%%%%%%%%%%%%%%%%%%%%%%%%%%%%%%%%%%%
%%%%%%%%%%%%%%%%%%%%section-end%%%%%%%%%%%%%%%%%%%%%%%%%%%
%%%%%%%%%%%%%%%%%%%%%%%%%%%%%%%%%%%%%%%%%%%%%%%%%%%%%%%%%%%%%%
%%%%%%%%%%%%%%%%%%%%%%%%%%%%%%%%%%%%%%%%%%%%%%%%%%%%%%%%%%%%%%

%%%%%%%%%%%%%%%%%%%%%%%%%%%%%%%%%%%%%%%%%%%%%%%%%%%%%%%%%%%%%%
%%%%%%%%%%%%%%%%%%%%%%%%%%%%%%%%%%%%%%%%%%%%%%%%%%%%%%%%%%%%%%
%%%%%%%%%%%%%%%%%%%%section-start%%%%%%%%%%%%%%%%%%%%%%%%%%%
%%%%%%%%%%%%%%%%%%%%%%%%%%%%%%%%%%%%%%%%%%%%%%%%%%%%%%%%%%%%%%
%%%%%%%%%%%%%%%%%%%%%%%%%%%%%%%%%%%%%%%%%%%%%%%%%%%%%%%%%%%%%%
\section{Case Studies}
\label{sec:casestudy}
This section evaluates the impact of the complete noise landscape on system performance. Three distance-based point-to-point scenarios are considered: short-range, medium-range, and long-range. All scenarios initially assume the Hot environmental condition defined in Table~\ref{tab:environmental_params} to establish a consistent baseline. An environmental sensitivity analysis then examines system robustness under Hot, Moderate, and Cold/Dry conditions.

\subsection{Short-Range Links}
\label{sec:shortrange}
%%%%%figure-start%%%%%%%
\begin{figure*}[!t]
    \centering
    \begin{subfigure}[b]{0.48\textwidth}
        \centering
        \includegraphics[width=\textwidth]{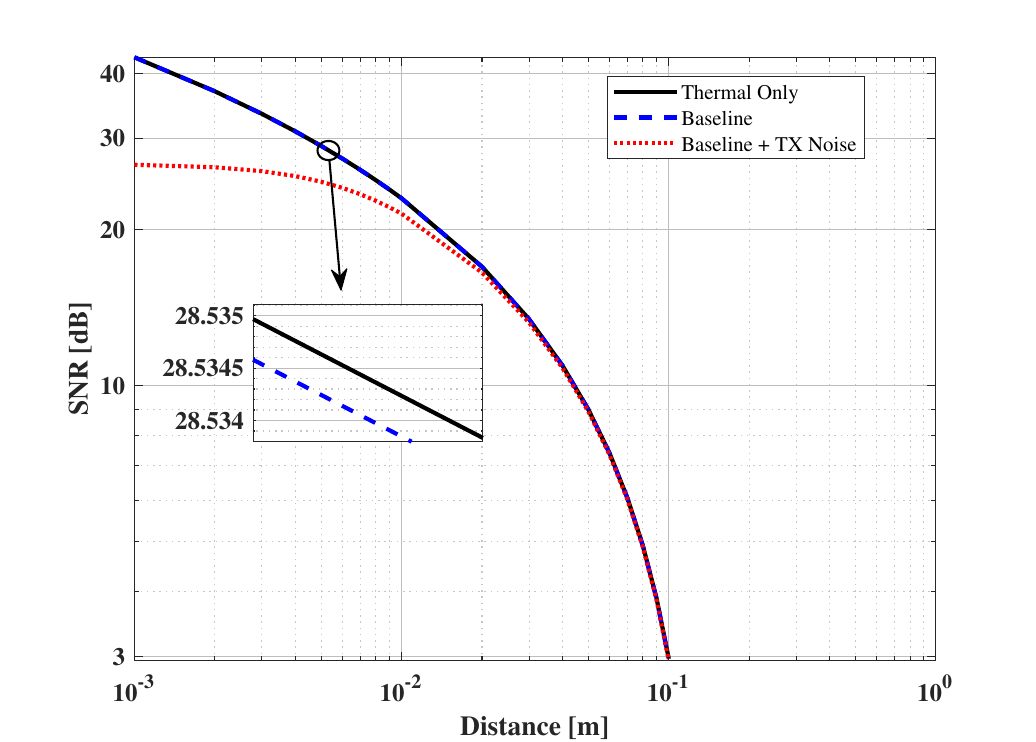}
        \caption{RX input SNR versus distance at 300 GHz (0 dBm input power) for different noise scenarios.}
        \label{fig:snr_vs_dist_short300}
    \end{subfigure}
    \hfill
    \begin{subfigure}[b]{0.48\textwidth}
        \centering
        \includegraphics[width=\textwidth]{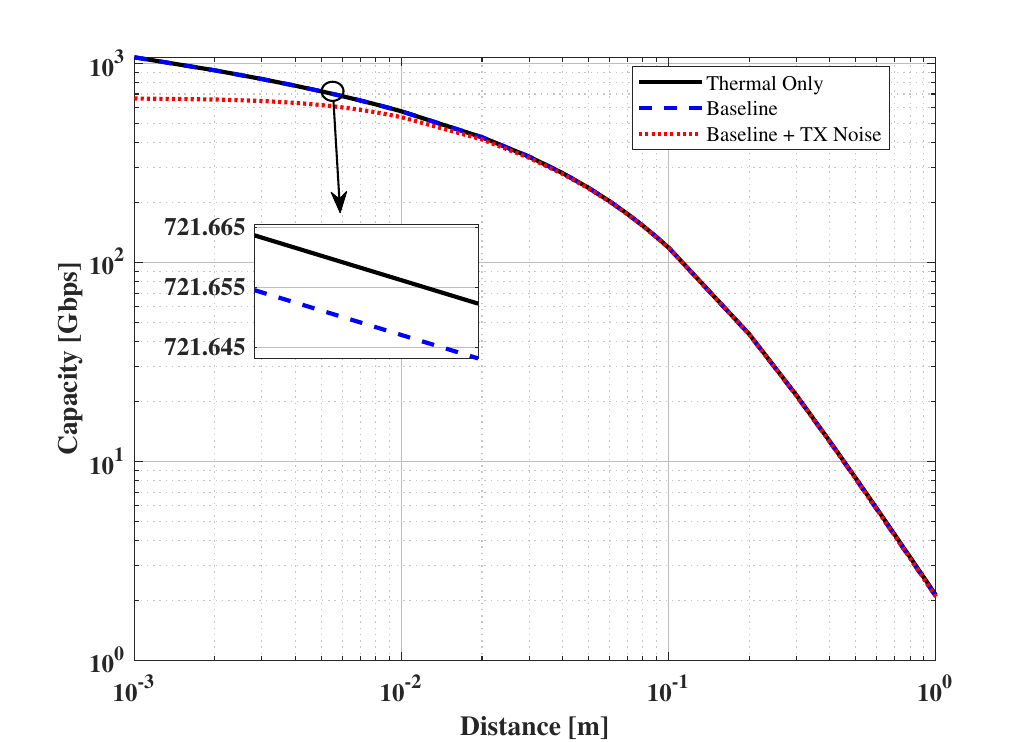}
        \caption{RX input capacity versus distance at 300 GHz (0 dBm input power) for different noise scenarios.}
        \label{fig:cap_vs_dist_short300}
    \end{subfigure}
    \caption{Short-range link performance at 300 GHz.}
    \label{fig:short_range_combined_300}
\end{figure*}
%%%%%figure-end%%%%%%%

%%%%%figure-start%%%%%%%
\begin{figure}[!t]
    \centerline{\includegraphics[width=\columnwidth]{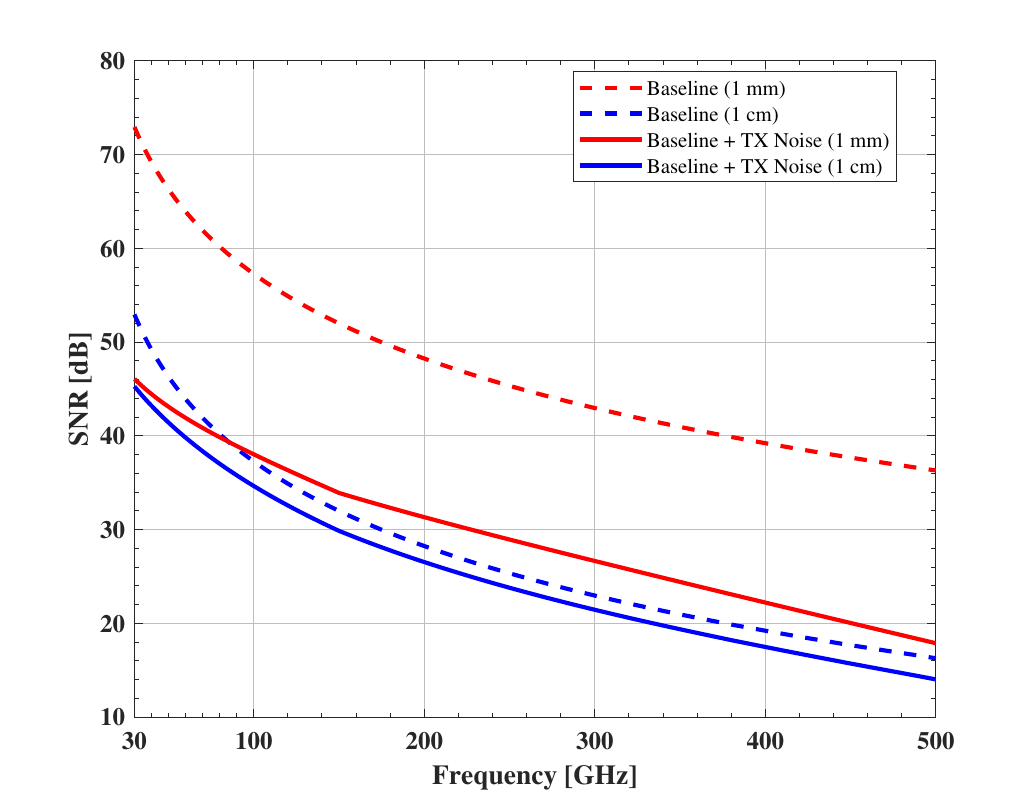}}
    \caption{SNR versus frequency at the RX input for short-range links (1~mm and 1~cm), comparing the baseline noise floor against the case where propagated TX noise is included.}
    \label{fig:short_snr_freq}
\end{figure}
%%%%%figure-end%%%%%%%

For short-range high data rate links (0.1--100 cm), such as those used in on-chip communication, device-to-device connectivity, and data center interconnects, our refined link budget model reveals important insights about noise contributions that significantly impact achievable performance. At these distances, atmospheric absorption effects are negligible due to the short propagation path. However, the correspondingly low total propagation loss is insufficient to attenuate TX noise below the RX's baseline noise floor, making propagated TX noise the primary performance limiting factor.

Figure~\ref{fig:snr_vs_dist_short300} illustrates the SNR versus distance for a
$300\,\mathrm{GHz}$ short-range link under different noise scenarios. At the shortest distance of $1\,\mathrm{mm}$, the Thermal Only and Baseline curves reach approximately $43\,\mathrm{dB}$ and are nearly indistinguishable, with atmospheric molecular noise contributing less than $0.001\,\mathrm{dB}$ of additional degradation. In contrast, the inclusion of propagated TX noise imposes a clear SNR ceiling at approximately $27\,\mathrm{dB}$, representing a reduction of about $16\,\mathrm{dB}$ relative to the baseline. As distance increases, this TX noise penalty decreases rapidly: at $1\,\mathrm{cm}$ the degradation is on the order of $2\,\mathrm{dB}$, and by $10\,\mathrm{cm}$ all three curves converge as the increasing propagation loss sufficiently attenuates the TX noise contribution.

Figure~\ref{fig:cap_vs_dist_short300} presents the corresponding channel capacity
versus distance. At $1\,\mathrm{mm}$ the thermal only capacity is slightly above
$720\,\mathrm{Gbps}$, the Baseline capacity is reduced by less than $0.02\,\mathrm{Gbps}$ relative to the Thermal Only case, confirming the negligible role of atmospheric molecular noise in the short-distance regime. Propagated TX noise, however, reduces the
maximum capacity to approximately $650\,\mathrm{Gbps}$, a reduction of about
$10\%$. As distance increases, the capacity gap narrows in the same manner as
the SNR curves, and beyond $10\,\mathrm{cm}$ all three scenarios converge due to
the dominance of propagation loss. At $300\,\mathrm{GHz}$, where the link budget
is extremely sensitive to SNR, even small penalties translate into reduced
spectral efficiency and lower feasible modulation orders.

The quantitative impact of propagated TX noise in short-range links is summarized in Table~\ref{tab:short_range_degradation}, which reveals two critical trends. First, the SNR degradation is extremely sensitive to distance. At 60~GHz, for example, the penalty drops from a massive 22~dB at 1~mm to approximately 4.1~dB at 10~mm, becoming negligible (0.06~dB) at 10~cm as path loss increases. Second, the impact is highly frequency dependent, with severe degradation occurring at lower frequencies (e.g., 22~dB at 60~GHz compared to 16.3~dB at 300~GHz). This is a direct result of the lower FSPL at lower frequencies, which provides insufficient attenuation of the propagated TX noise. The data confirms that for all analyzed frequencies, the TX noise degradation becomes minimal ($<$0.1~dB) by 10~cm and is effectively zero by 1~m.

To provide additional physical insight into the distance- and frequency-dependent trends summarized in Table~\ref{tab:short_range_degradation}, Fig.~\ref{fig:short_snr_freq} depicts the frequency-domain SNR at the RX input for fixed short-range separations of 1~mm and 1~cm. The figure compares the Baseline noise floor against the case where propagated TX noise is included. At a separation of 1~mm, the inclusion of TX noise introduces an SNR degradation of approximately 27~dB at 30~GHz. As frequency increases, this impact decreases, reaching about 16.5~dB at 300~GHz and 18.4~dB at 500~GHz. When the distance increases to 1~cm, the impact of TX noise is significantly reduced due to the increased FSPL. For example, at 30~GHz, TX noise introduces approximately 7~dB of additional degradation relative to the Baseline, while this penalty decreases to about 1.5~dB at 300~GHz.

The implications of this analysis for short-range system design are significant. The results demonstrate that propagated TX noise, not thermal noise, is the primary factor limiting performance, creating a fundamental SNR ceiling of approximately 27 dB. This ceiling directly restricts achievable channel capacity and limits practical modulation orders. Consequently, for short-range sub-THz systems, design efforts must prioritize reducing the TX's cascaded noise figure, as the traditional thermal noise only approach leads to systematic overestimation of achievable performance.

%%%%%table-start%%%%%%%
\begin{table}[!t]
\centering
\caption{TX Noise Impact on SNR at Short Ranges}
\label{tab:short_range_degradation}
\begin{tabular}{@{}lcccc@{}}
\toprule
\textbf{Distance} & \multicolumn{4}{c}{\textbf{SNR Degradation [dB]}} \\
\cmidrule(l){2-5}
& \textbf{60 GHz} & \textbf{140 GHz} & \textbf{250 GHz} & \textbf{300 GHz} \\
\midrule
1 mm   & 22 & 18.1 & 16.4 & 16.3 \\
10 mm  & 4.1 & 2.1  & 1.5  & 1.5  \\
100 mm & 0.06 & 0.02  & 0.01 & 0.01 \\
1 m    & 0.0006 & N/A† & N/A† & N/A† \\
\bottomrule
\end{tabular}
\vspace{0.3cm}
\footnotesize
\par
SNR degradation = SNR$_{\text{Baseline}}$ (Thermal+Atmospheric) - SNR$_{\text{All}}$ (Thermal+Atmospheric+TX)
\par
†Link not viable (negative SNR at baseline)
\end{table}
%%%%%table-end%%%%%%%

\subsection{Medium-Range Links}
\label{sec:mediumrange}
%%%%%figure-start%%%%%%%
\begin{figure*}[!t]
    \centering
    \begin{subfigure}[b]{0.48\textwidth}
        \centering
        \includegraphics[width=\textwidth]{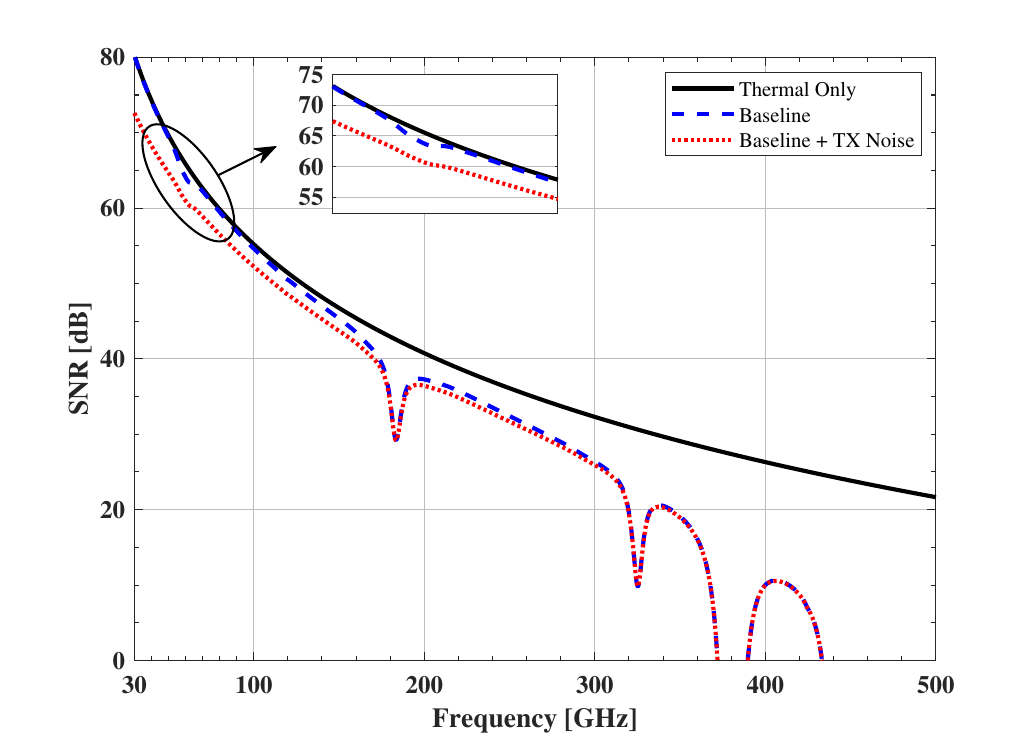}
        \caption{SNR versus frequency at RX input for 100 m distance.}
        \label{fig:mediumsnr_vs_freq}
    \end{subfigure}
    \hfill
    \begin{subfigure}[b]{0.48\textwidth}
        \centering
        \includegraphics[width=\textwidth]{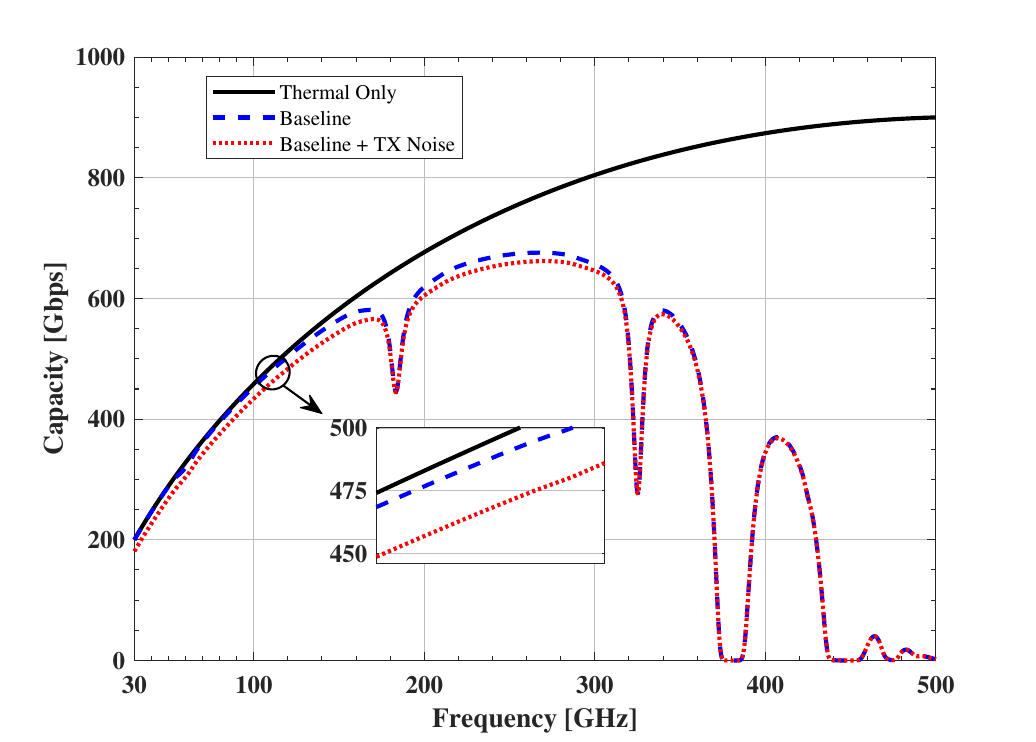}
        \caption{Capacity versus frequency at RX input for 100 m distance.}
        \label{fig:mediumcap_vs_freq}
    \end{subfigure}
    \caption{Medium-range link performance at 100 m.}
    \label{fig:medium_range_combined}
\end{figure*}
%%%%%figure-end%%%%%%%

%%%%%figure-start%%%%%%%
\begin{figure*}[!t]
    \centering
    \begin{subfigure}[b]{0.48\textwidth}
        \centering
        \includegraphics[width=\textwidth]{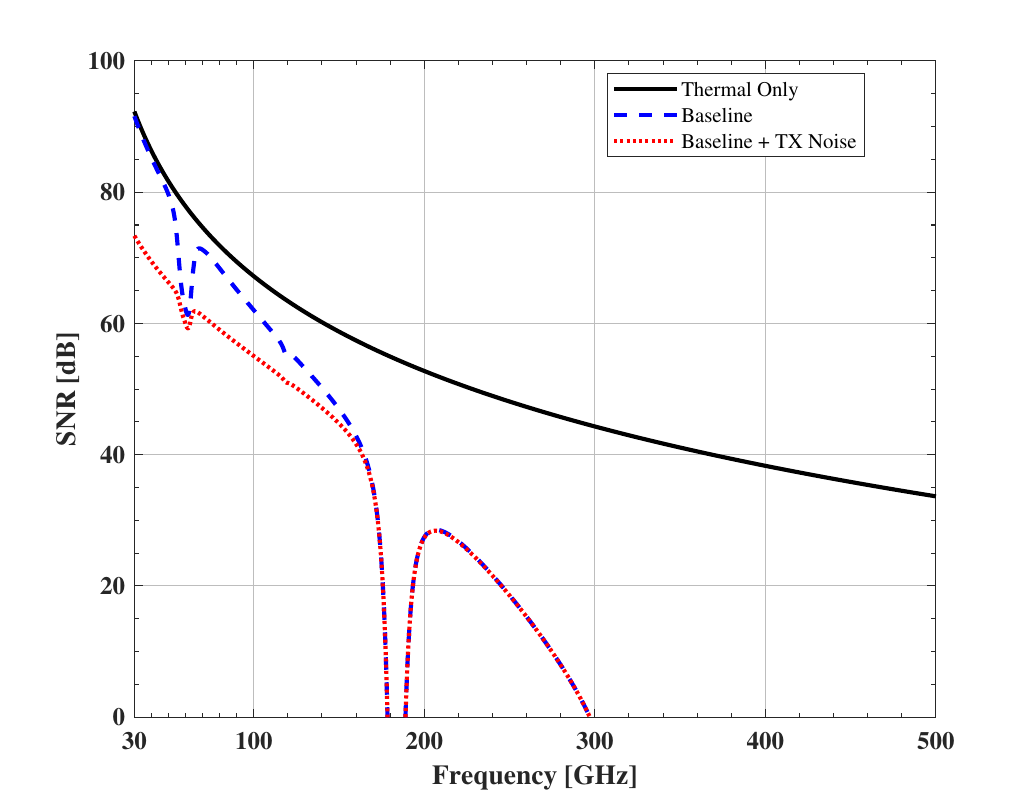}
        \caption{SNR versus frequency at RX input for 1000 m distance.}
        \label{fig:longsnr_vs_freq}
    \end{subfigure}
    \hfill
    \begin{subfigure}[b]{0.48\textwidth}
        \centering
        \includegraphics[width=\textwidth]{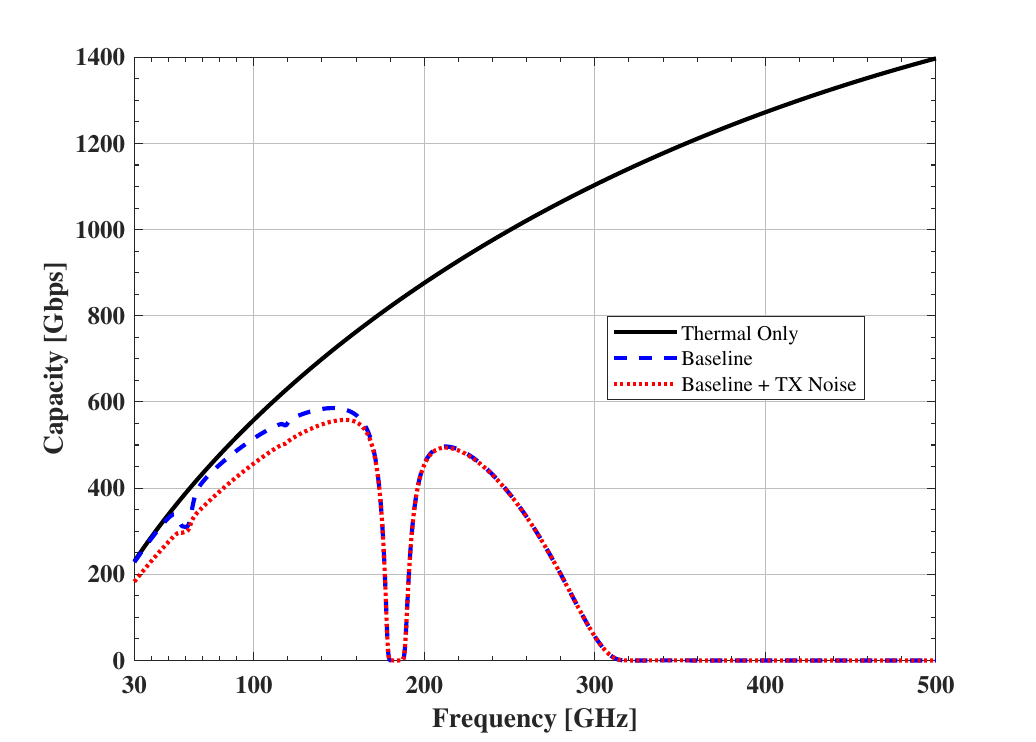}
        \caption{Channel capacity versus frequency at RX input for 1000 m distance.}
        \label{fig:longcap_vs_freq}
    \end{subfigure}
    \caption{Long-range link performance at 1000 m.}
    \label{fig:long_range_combined}
\end{figure*}
%%%%%figure-end%%%%%%%

Medium-range links operate at distances on the order of 100~m. At these separations, free-space path loss and atmospheric molecular absorption become significant, while propagated TX noise remains non-negligible. As a result, medium-range operation represents a transition regime in which the Thermal Only, Baseline, and Baseline + TX Noise contributions all influence link performance. Comprehensive noise modeling is therefore essential for accurate performance prediction in this distance range.

Figure~\ref{fig:mediumsnr_vs_freq} illustrates the SNR versus frequency for a $100\,\mathrm{m}$ link. The Thermal Only curve decreases smoothly from approximately $82\,\mathrm{dB}$ at $30\,\mathrm{GHz}$ to $22\,\mathrm{dB}$ at $500\,\mathrm{GHz}$. The Baseline introduces pronounced frequency-selective degradation due to atmospheric molecular absorption, with absorption notches reducing the SNR to approximately $15\,\mathrm{dB}$ at $183\,\mathrm{GHz}$ and about $20\,\mathrm{dB}$ at $325\,\mathrm{GHz}$. Even outside these absorption peaks, atmospheric effects reduce the SNR by approximately $5$ to $10\,\mathrm{dB}$ relative to the Thermal Only case, as highlighted in the inset. The inclusion of propagated TX noise (Baseline + TX Noise) produces additional degradation, most clearly visible in the $30$--$110\,\mathrm{GHz}$ range where it introduces a $2$ to $8\,\mathrm{dB}$ penalty. At intermediate frequencies ($100$--$200\,\mathrm{GHz}$), this penalty decreases to approximately $1\,\mathrm{dB}$ as atmospheric losses increasingly dominate the total noise floor.

Figure~\ref{fig:mediumcap_vs_freq} presents the corresponding channel capacity. The Thermal Only capacity exceeds $900\,\mathrm{Gbps}$ near $500\,\mathrm{GHz}$, while atmospheric absorption limits the Baseline capacity to approximately $650$--$700\,\mathrm{Gbps}$ within the clear spectral windows, with deep nulls at $60$, $183$, and $325\,\mathrm{GHz}$. The inclusion of propagated TX noise (Baseline + TX Noise) introduces an additional capacity reduction of roughly $10$--$25\,\mathrm{Gbps}$ in the lower frequency region ($30$--$80\,\mathrm{GHz}$), as emphasized in the inset. These results indicate that for medium-range links, atmospheric molecular absorption determines the viable operating windows, while TX noise imposes a secondary but consistent performance penalty at the lower end of the spectrum.

\subsection{Long Range Links}
\label{sec:longrange}
Long-range links operate at distances of 1000~m and beyond, where extremely high total propagation loss fundamentally changes the noise landscape. The severe attenuation, often exceeding 140~dB across many frequency bands, effectively suppresses propagated TX noise, rendering it negligible. Consequently, system performance becomes dominated by the combined effects of RX thermal noise and atmospheric molecular noise.

Figure~\ref{fig:longsnr_vs_freq} illustrates the SNR versus frequency for a $1000\,\mathrm{m}$ long-range link. Atmospheric molecular absorption dominates the overall behavior, producing notches at the $60\,\mathrm{GHz}$ oxygen band and the $183\,\mathrm{GHz}$ and $325\,\mathrm{GHz}$ water vapor bands, where the SNR collapses to nearly zero. Even away from these absorption peaks, the Baseline reduces the SNR by approximately $10$ to $15\,\mathrm{dB}$ relative to the Thermal Only case across much of the $30$--$200\,\mathrm{GHz}$ range. The inclusion of propagated TX noise (Baseline + TX Noise) introduces additional degradation that is most pronounced at the lower end of the spectrum. In the $30$--$80\,\mathrm{GHz}$ region, TX noise reduces the SNR by approximately $5$ to $15\,\mathrm{dB}$ relative to the baseline. As frequency increases, the impact of TX noise diminishes, and in the $100$--$200\,\mathrm{GHz}$ range the baseline and Baseline + TX Noise curves nearly converge. Beyond $200\,\mathrm{GHz}$, atmospheric absorption becomes sufficiently severe that the contribution of TX noise is negligible by comparison. These results indicate that for long-range operation, atmospheric molecular absorption is the dominant limiting factor, while TX noise introduces a secondary penalty primarily at lower frequencies.

Figure~\ref{fig:longcap_vs_freq} shows the corresponding channel capacity for the $1000\,\mathrm{m}$ link. The Thermal Only case exceeds $1300\,\mathrm{Gbps}$ near $500\,\mathrm{GHz}$; however, atmospheric absorption severely limits the usable spectrum. Within the clear transmission windows, the practical capacity is constrained to approximately $500$--$650\,\mathrm{Gbps}$, while the absorption notches at $183$ and $325\,\mathrm{GHz}$ result in nearly zero capacity. The inclusion of propagated TX noise (Baseline + TX Noise) reduces capacity primarily at the lower end of the frequency range. Between $30$ and $110\,\mathrm{GHz}$, TX noise lowers capacity by approximately $30$ to $60\,\mathrm{Gbps}$ relative to the baseline. At higher frequencies, this difference becomes small as atmospheric absorption dominates the total noise power. Consistent with the SNR results, long-range links are therefore fundamentally limited by the atmospheric absorption profile, with TX noise introducing an additional but frequency-dependent performance penalty.

%%%%%table-start%%%%%%%
\begin{table*}[!t]
    \centering
    \caption{Guidance on Dominant Noise Source Considerations in mm-Wave and sub-THz Link Budgets}
    \label{tab:noise_conclusion_summary}
    \renewcommand{\arraystretch}{1.4}
    \begin{tabular}{@{} l l m{4.2cm} m{4.2cm} @{}}
        \toprule
        \textbf{Frequency Band} & \textbf{Distance Category} & \textbf{Atmospheric Molecular Noise} & \textbf{TX Noise} \\
        \midrule
        % 30-100 GHz Section
        \multirow{3}{*}{\parbox{2.5cm}{\centering\textbf{30-100 GHz}}} 
        & Short Range ($<$1~m) & Low  & Critical \\
        \cmidrule(l){2-4}
        & Medium Range (100m) & Critical & Moderate  \\
        \cmidrule(l){2-4}
        & Long Range (1km) & Critical & Significant  \\
        \midrule
        % 100-300 GHz Section
        \multirow{3}{*}{\parbox{2.5cm}{\centering\textbf{100-200 GHz}}}
        & Short Range ($<$1~m) & Low  & Critical  \\
        \cmidrule(l){2-4}
        & Medium Range (100m) & Critical & Moderate \\
        \cmidrule(l){2-4}
        & Long Range (1km) & Critical  & Low \\
        \midrule
        % 300-500 GHz Section
        \multirow{3}{*}{\parbox{2.5cm}{\centering\textbf{200-500 GHz}}}
        & Short Range ($<$1~m) & Low  & Critical  \\
        \cmidrule(l){2-4}
        & Medium Range (100m) & Critical  & Low  \\
        \cmidrule(l){2-4}
        & Long Range (1km) & Critical & Low  \\
        \bottomrule
    \end{tabular}
    \par\medskip
    \small{ \textit{Low:} $<$1 dB impact; \textit{Moderate:} 1-3 dB; \textit{Significant:} 3-5 dB; \textit{Critical:} $>$5 dB or dominant noise source. Thresholds reflect stringent link budget constraints in mm-Wave and sub-THz systems. }
\end{table*}
%%%%%table-end%%%%%%%

The preceding case studies demonstrate that the dominant noise source in mm-Wave and sub-THz systems is highly dependent on the link distance, as consolidated in Table~\ref{tab:noise_conclusion_summary}. For short-range links, propagated TX noise emerges as the critical impairment, creating performance ceilings that are only weakly dependent on path loss. Medium-range links represent a transition regime in which atmospheric molecular noise becomes a major limiting factor, while propagated TX noise continues to impose a non-negligible performance penalty. For long-range links, atmospheric noise dominates frequency selection and overall link viability; however, propagated TX noise remains a relevant impairment in the lower-frequency atmospheric transmission windows (e.g., 30--110~GHz). These results confirm that a comprehensive, multi-source noise model is essential for accurate system design across all mm-Wave and sub-THz link scenarios.

%%%%%figure-start%%%%%%%
\begin{figure}[!t]
    \centerline{\includegraphics[width=\columnwidth]{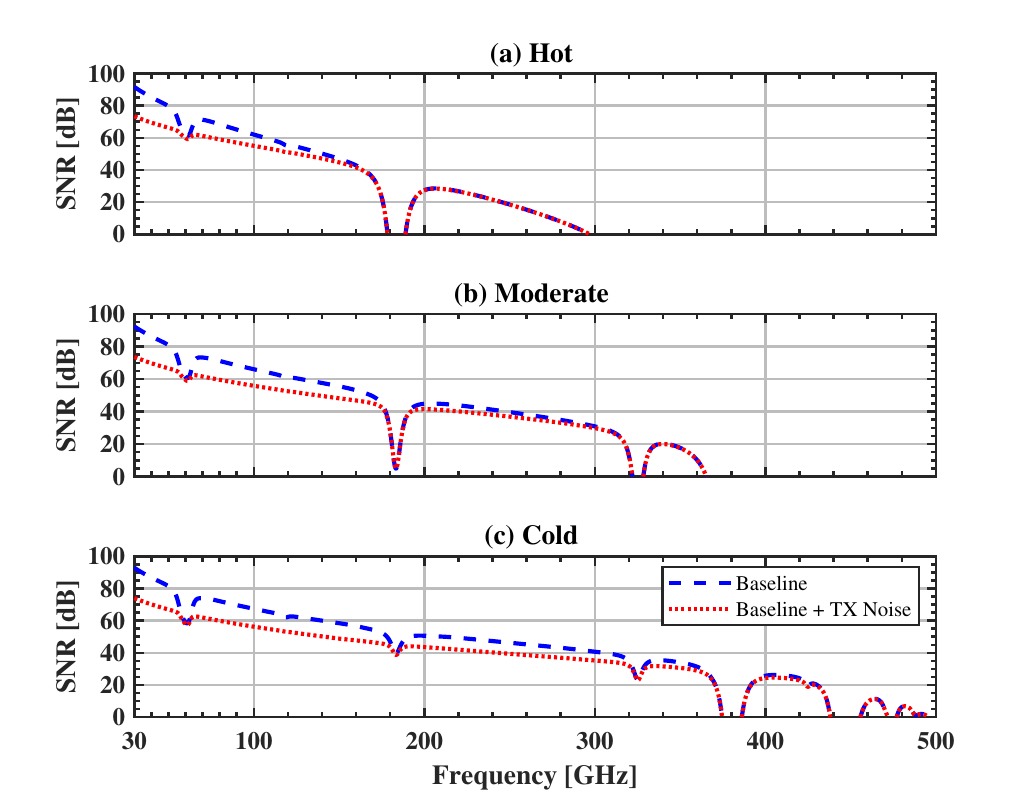}}
    \caption{Environmental sensitivity analysis showing SNR versus frequency at the 1000m range under different atmospheric conditions.}
    \label{fig:weather}
\end{figure}
%%%%%figure-end%%%%%%%

\subsection{Environmental Sensitivity Analysis}
\label{sec:environmental_sensitivity}
To evaluate system robustness under varying environmental conditions, we analyze three representative scenarios: Hot, Moderate, and Cold/Dry, with parameters listed in Table~\ref{tab:environmental_params}. These parameters follow established atmospheric models~\cite{usstandard1976} for pressure and temperature baselines, with corresponding water vapor densities derived from standard temperature-humidity relations~\cite{wmo}.

Figure~\ref{fig:weather} illustrates the environmental sensitivity of long-range links by comparing the SNR versus frequency at a distance of $1000\,\mathrm{m}$ under Hot, Moderate, and Cold/Dry atmospheric conditions. In the Hot condition, atmospheric molecular absorption is strongest, producing deep notches at the $183\,\mathrm{GHz}$ and $325\,\mathrm{GHz}$ water vapor lines where the SNR collapses to nearly zero. In this regime, atmospheric noise dominates the total noise floor, and the impact of propagated TX noise is largely confined to the lower-frequency region below approximately $120\,\mathrm{GHz}$.

The Moderate condition exhibits weaker atmospheric absorption than the Hot case, leading to an overall SNR increase of several decibels outside the absorption peaks. Under these conditions, the contribution of propagated TX noise becomes more apparent, particularly in the $50$--$150\,\mathrm{GHz}$ range where atmospheric attenuation is reduced and the difference between the Baseline and Baseline + TX Noise cases reaches several decibels. The Cold/Dry condition produces the weakest atmospheric absorption due to reduced water vapor content. In this environment, the SNR outside the absorption lines is significantly higher, which exposes the contribution of propagated TX noise across a broad portion of the spectrum. In particular, between $50$ and $250\,\mathrm{GHz}$, the TX noise penalty is on the order of $3$ to $6\,\mathrm{dB}$, indicating that TX hardware quality becomes a primary performance limitation when atmospheric molecular noise is suppressed. These results demonstrate that the relative importance of propagated TX noise increases substantially under dry atmospheric conditions.

To directly guide hardware and system designers in prioritizing noise considerations, Table~\ref{tab:noise_conclusion_summary} synthesizes the findings across the complete 30--500~GHz spectrum. The table highlights when each noise source becomes a dominant performance factor and when its impact is secondary across different frequency bands and link distances. This mapping of the noise landscape enables designers to allocate component specifications and link budget margins more effectively, prioritizing TX noise figure reduction for short-range systems while emphasizing atmospheric window selection for long-range deployments.

%%%%%%%%%%%%%%%%%%%%%%%%%%%%%%%%%%%%%%%%%%%%%%%%%%%%%%%%%%%%%%
%%%%%%%%%%%%%%%%%%%%%%%%%%%%%%%%%%%%%%%%%%%%%%%%%%%%%%%%%%%%%%
%%%%%%%%%%%%%%%%%%%%section-end%%%%%%%%%%%%%%%%%%%%%%%%%%%
%%%%%%%%%%%%%%%%%%%%%%%%%%%%%%%%%%%%%%%%%%%%%%%%%%%%%%%%%%%%%%
%%%%%%%%%%%%%%%%%%%%%%%%%%%%%%%%%%%%%%%%%%%%%%%%%%%%%%%%%%%%%%

%%%%%%%%%%%%%%%%%%%%%%%%%%%%%%%%%%%%%%%%%%%%%%%%%%%%%%%%%%%%%%
%%%%%%%%%%%%%%%%%%%%%%%%%%%%%%%%%%%%%%%%%%%%%%%%%%%%%%%%%%%%%%
%%%%%%%%%%%%%%%%%%%%section-start%%%%%%%%%%%%%%%%%%%%%%%%%%%
%%%%%%%%%%%%%%%%%%%%%%%%%%%%%%%%%%%%%%%%%%%%%%%%%%%%%%%%%%%%%%
%%%%%%%%%%%%%%%%%%%%%%%%%%%%%%%%%%%%%%%%%%%%%%%%%%%%%%%%%%%%%%
\section{CONCLUSION}
This paper has presented a comprehensive link budget framework for mm-Wave and sub-THz systems that explicitly incorporates both propagated TX noise and atmospheric molecular noise. An analytical criterion was derived to determine the conditions under which TX noise dominates system performance, revealing a strong dependence on frequency and link distance. At short ranges (below $1\,\mathrm{m}$), propagated TX noise imposes fundamental SNR ceilings, producing degradations on the order of $15$ to $25\,\mathrm{dB}$ depending on operating frequency and hardware quality. At medium ($100\,\mathrm{m}$) and long ($1\,\mathrm{km}$) ranges, atmospheric molecular noise becomes the prevailing factor, while the influence of propagated TX noise diminishes and remains relevant primarily within the lower-frequency atmospheric transmission windows. The results of this study provide practical guidance for optimizing component selection and link configuration across the $30$--$500\,\mathrm{GHz}$ spectrum.

%\section*{APPENDIX}
%Appendixes, if needed, appear before the acknowledgment.

%\section*{ACKNOWLEDGMENT}
%This work was financially supported by TUBITAK 121C073 and Horizon Europe MSCA Post-Doctoral Fellowship (101062272-ENSPEC6G).


%% ===============================
%% BIOGRAPHIES (removed for arXiv preprint)
%% ===============================

\end{document}